\documentclass[journal=mamobx,manuscript=article]{achemso}

\usepackage{geometry}
\usepackage{lineno,hyperref,amsmath,amssymb, subcaption, bm, multirow}
\usepackage[thinlines]{easytable}

\DeclareMathOperator{\Tr}{tr\hspace{1pt}}
\newcommand{\h}[1]{\hat{#1}}

\usepackage{setspace}
\usepackage{enumitem} 
\usepackage{siunitx}
\DeclareSIUnit\da{\text{Da}}
\DeclareSIUnit\wtpc{\text{wt\%}}

\usepackage[referable]{threeparttablex}
\renewlist{tablenotes}{enumerate}{1}
\makeatletter
\setlist[tablenotes]{label=\tnote{\alph*},ref=\alph*,itemsep=\z@,topsep=\z@skip,partopsep=\z@skip,parsep=\z@,itemindent=\z@,labelindent=\tabcolsep,labelsep=.2em,leftmargin=*,align=left,before={\footnotesize}}
\makeatother

\newcommand{\mr}[1]{\mathrm{#1}}

\usepackage{tikz}
\newcommand\drawlinesc[1][solid]{%
  \begin{tikzpicture}
    \draw[#1] (0,0) -- (5mm,0);                                       
  \end{tikzpicture}%
} 
\definecolor{sol1}{RGB}{62, 53, 107}
\definecolor{sol2}{RGB}{53, 123, 163}
\definecolor{sol3}{RGB}{75, 194, 173}
\definecolor{sol4}{RGB}{254, 227, 144}
\definecolor{sol5}{RGB}{254, 152, 41}
\definecolor{sol6}{RGB}{203, 75, 2}
    
\usepackage{layouts}
\usepackage{booktabs}
\usepackage{multirow}

\author{Jianyi Du}
\affiliation[Massachusetts Institute of Technology]
{Hatsopoulos Microfluids Laboratory, Department of Mechanical Engineering, Massachusetts Institute of Technology, Cambridge, MA 02139, United States}
\author{Hiroko Ohtani}
\affiliation[Ford Motor Company]{Ford Research and Innovation Center, Dearborn, MI 48124, United States}
\author{Alper Kiziltas}
\affiliation[Ford Motor Company]{Ford Research and Innovation Center, Dearborn, MI 48124, United States}
\author{Kevin Ellwood}
\affiliation[Ford Motor Company]{Ford Research and Innovation Center, Dearborn, MI 48124, United States}
\author{Gareth H. McKinley}
\affiliation[Massacbhusetts Institute of Technology]
{Hatsopoulos Microfluids Laboratory, Department of Mechanical Engineering, Massachusetts Institute of Technology, Cambridge, MA 02139, United States}
\email{gareth@mit.edu}

\title[rpcaber]{Capillarity-driven thinning dynamics of entangled polymer solutions}


\begin{document}


\begin{abstract}
We analyze the capillarity-driven thinning dynamics of entangled polymer solutions described by the Doi-Edwards-Marrucci-Grizzuti (DEMG) model and the Rolie-Poly (RP) model. Both models capture polymer reptation, finite rates of chain retraction and finite extensibility of single polymer molecules, while differing slightly in their final form regarding to the convective constraint release. We calculate numerically the filament thinning profiles predicted by the two models with realistic entanglement densities, assuming cylindrical filament shapes and no fluid inertia. Both results reveal an early tube-reorientation regime, followed by a brief intermediate elasto-capillary regime, and finally a finite-extensibility regime close to the pinch-off singularity. The results presented in this work reveal two critical features in the transient extensional rheology of entangled polymer solutions that have been reported from previous experimental studies, but are poorly described by the widely-used FENE-P model. First, the relaxation time obtained from capillary breakup extensional rheometry is notably smaller than that from steady-shear rheometry. Their ratio can be expressed as a universal function of the entanglement state and the polymer concentration, which agrees well with the experimental data for a range of entangled polymer solutions. Second, the filament thinning dynamics at sufficiently high polymer concentrations are governed by the tube reorientation at intermediate strain-rates, and the apparent extensional viscosity shows a noticeably rate-thinning response. We finally evaluate the filament thinning dynamics of aqueous polyethylene oxide solutions (\SI{1}{\mega\da}) over dilute and entangled regimes. As the concentration increases, the profiles deviate from the well-studied exponential-thinning trends beyond the entangled threshold, becoming increasingly power-law in character.
\end{abstract}



\section{Introduction}
It is well known that the extensional rheological behavior of entangled polymer solutions or polymer melts exhibit distinct properties compared with those of dilute polymer solutions \cite{larsonModelingRheologyPolymer2015}. In these entangled systems, strong intermolecular interactions result in more complex microstructural dynamics that lead to distinct relaxation mechanisms over a wide range of time- and lengthscales during material deformation. Consequently, most entangled polymer systems exhibit strongly nonlinear rheological properties, such as rate-thinning viscosities, stress overshoots following step-strains and onset of transient elastic instabilities \cite{larsonModelingRheologyPolymer2015,dealyStructureRheologyMolten2018}. Numerous industrial processes depend on understanding the non-linear extensional response of these materials, including polymer fiber spinning, food processing, injection or plastic blow molding and filament extrusion \cite{mckinleyFILAMENTSTRETCHINGRHEOMETRYCOMPLEX2002}.

Over the past few decades, extensive efforts have been made to characterize and model the rheological complexity of entangled polymer systems \cite{larsonModelingRheologyPolymer2015}. Significant progress has been made since the milestone work of reptation theory, which provides a coarse-grained canonical framework to understand the dynamical response of entangled polymer chains \cite{narimissaReviewTubeModel2019}. Each reptating polymer chain is envisioned to be fully or partially confined by surrounding polymer chains in a mean-field imaginary tube. The entangled chains interact with each other at their topological crossovers, or entanglements, and these geometrical constraints inhibit the transverse motion of a single chain and only allow for diffusive motion, or ``reptation,'' along the imaginary tube with a greatly reduced diffusivity \cite{degennesReptationPolymerChain1971, hiemenz2007polymer}. This low diffusivity in the tube results in a significant slowdown in the dominant polymer relaxation dynamics. Based on this concept, Doi and Edwards \cite{doi1978dynamics} proposed a full-dimensional constitutive model for monodisperse entangled linear polymers derived from the Lodge rubberlike liquid \cite{birdDynamicsPolymericLiquids1987a}. In the original Doi-Edwards (DE) model, the tube segments are reoriented in the flow direction based on the ``independent alignment approximation (IAA).'' Under the IAA approximation, the stress originates from the affine transformation of the tube segments instead of the tube stretch. Because of the extensibility of tube segment is not considered, the primitive chains between entanglements are simplified as rigid rods. The IAA only applies to an intermediate strain-rate range of $1/\lambda_\mr{D}\lesssim \dot{\gamma} \ll 1/\lambda_\mr{R}$ in the stress relaxation process \cite{larsonModelingRheologyPolymer2015}. Here, $\lambda_\mr{R}$ is the Rouse time quantifying the time for retraction of a single polymer chain in the tube, and $\lambda_\mr{D}$ is the disengagement time characterizing the timescale over which a single polymer chain reptates out of its constraining tube.
These two timescales can be connected through the relationship \cite{robertsonTheoreticalPredictionExperimental2017,
likhtmanQuantitativeTheoryLinear2002} as
\begin{subequations}
\begin{align}
\lambda_\mr{D}&=p(Z)\lambda_\mr{R}=p(Z)\dfrac{\xi_0 (Nb)^2}{6 \pi^2 k T},\\
p(Z)&=3Z(1-\dfrac{3.38}{\sqrt{Z}}+\dfrac{4.17}{Z}-\dfrac{1.55}{Z^{3/2}}),
\end{align}
\label{eqn:Z}%
\end{subequations}
where the Rouse time $\lambda_\mr{R}$ is explicitly defined as a function of the monomeric friction coefficient $\xi_0$ and the statistical contour length $Nb$ \cite{dealyMolecularStructureRheology2006}. $Z=M/M_\mr{e}$ is the number of entanglements per polymer chain \cite{dealyStructureRheologyMolten2018}, and $M$ and $M_\mr{e}$ are the molecular weights of a single polymer chain and one entangled tube segment, respectively. The form of $p(Z)$ is a truncated Taylor expansion of a function of $Z^{-1/2}$ numerically calculated by Likhtman and McLeish \cite{likhtmanQuantitativeTheoryLinear2002} to incorporate the effects of contour length fluctuation (CLF) in polymer chains of realistic lengths and numbers of entanglements.

The DE model successfully captures a number of key rheological features of an entangled polymer system, such as a molecular-weight independent storage modulus, a non-trivial second-order normal stress difference in steady shear flow and strain-softening behavior in a step-strain experiment at large strains \cite{oberhauserNonNewtonianFluidMechanics2000,dealyStructureRheologyMolten2018}. However, this model is well-known to underpredict the scaling of the zero-shear viscosity with the molecular weight, giving $\eta_0\propto M^3$, compared with $\eta_0\propto M^{3.4}$ from experiments. The DE model also predicts excessive levels of shear- and extension-thinning in steady shear and extensional flows respectively, leading to a strong susceptibility to flow instabilities \cite{dealyStructureRheologyMolten2018,narimissaReviewTubeModel2019}. Subsequent studies have attributed these deficiencies in the DE model to the absence of two non-reptative contributions to the chain dynamics: (1) a finite rate of chain retraction on short timescales or high strain rates (\textit{i.e.}, $t<\lambda_\mr{R}$ or $\lambda_\mr{R}\dot{\gamma}>1$), and (2) convective constraint release on longer timescales of a relaxation process (\textit{i.e.}, $t>\lambda_\mr{D}$ or $1/\lambda_\mr{D}<\dot{\gamma}<1/\lambda_\mr{R}$) \cite{marrucci1983free}. These deficiencies have subsequently been accounted for in a number of more sophisticated constitutive models based on the underlying micro-mechanical framework of reptation, including the Doi-Edwards-Marrucci-Grizzuti model \cite{marrucci1983free}, the GLaMM model \cite{grahamMicroscopicTheoryLinear2003} and the Rolie-Poly model \cite{likhtmanSimpleConstitutiveEquation2003}. 

Over the past twenty years, the capillarity-driven thinning technique has provided an accurate and efficient protocol to probe the transient extensional rheological properties of complex fluids. In this technique, a small initially cylindrical fluid sample is placed between two coaxial discs and then stretched by a fast axial strain. The resulting deformation - driven by the Rayleigh-Plateau instability - produces a thin, elongated filament, which subsequently undergoes a transient self-thinning process through the action of capillarity. When inertial and gravitational effects are negligible, the temporal evolution of the filament profile, resulting from the balance between the capillary pressure and the viscoelastic response from the fluid, can be used to extract a measure of the transient extensional rheological properties of the material \cite{mckinleyViscoelastocapillaryThinningBreakup2005}. A number of variations of the capillarity-driven thinning technique are now available. Examples include the Capillary Breakup Extensional Rheometer (CaBER) \cite{mckinleyViscoelastocapillaryThinningBreakup2005}, the Rayleigh-Ohnesorge Jetting Extensional Rheometer (ROJER) \cite{keshavarzStudyingEffectsElongational2015} and the Dripping-On-Substrate (DOS) method \cite{dinicExtensionalRelaxationTimes2015}. These techniques enable determination of the extensional rheological properties for a number of complex material systems, including Newtonian fluids \cite{mckinleyHowExtractNewtonian2000}, dilute polymer solutions \cite{entovEffectSpectrumRelaxation1997, clasenHowDiluteAre2006, tirtaatmadja2006drop, duImprovedCapillaryBreakup2021}, yield-stress fluids and emulsions \cite{niedzwiedzExtensionalRheologyConcentrated2010,martinieApparentElongationalYield2013, moschopoulosDynamicsViscoplasticFilament2020}, particulate suspensions \cite{tiwariElongationalShearRheology2009,
mcilroyModellingCapillaryBreakup2014,
ngGOCaBERCapillary2020}, magnetorheological fluids \cite{sadekCapillaryBreakupExtensional2019}, as well as more complex systems such as cellulose solutions \cite{hawardShearExtensionalRheology2012}, liquid food additives \cite{torresEffectConcentrationShear2014, jimenezCapillaryBreakupExtensional2020a} and cosmetic products such as nail varnishes \cite{jimenezRheologicallycomplexFluidBeauty2021}.

However, an explicit calculation of the filament thinning dynamics for entangled polymer systems has not been presented. A handful of studies \cite{arnoldsCapillaryBreakupExtensional2010,
sachsenheimerExperimentalStudyCapillary2014,
sachsenheimer2014elongational, dinicPowerLawsDominate2020}
have experimentally characterized the capillarity-driven thinning dynamics of aqueous polyethylene oxide (PEO), polystyrene (PS), hydroxy-ethyl cellulose (HEC), and wormlike micelle solutions across the dilute and entangled regimes. As the concentration increases beyond the entanglement threshold, the temporal evolution of the mid-filament radius measured experimentally deviates notably from the well-studied exponential thinning result described by an elasto-capillary balance of the Hookean dumbbell model \cite{entovEffectSpectrumRelaxation1997,birdDynamicsPolymericLiquids1987}. Observations reveal two unique features in the transient extensional rheological behavior of the entangled regime. First, the apparent extensional relaxation time $\lambda_\mr{e}$ (fitted from the exponential-thinning evolution of the filament radius) is evidently smaller than that independently extracted from shear flow (which we denote here as $\lambda_\mr{s}$), and the ratio of the two relaxation times $\lambda_\mr{e}/\lambda_\mr{s}$ determined in shear and extension is found to decrease at higher concentrations. Arnold et al. \cite{arnoldsCapillaryBreakupExtensional2010} have suggested the use of a damping function of the K-BKZ form to reconcile the different relaxation times in shear and extensional flows. However, additional fitting parameters are introduced to the resulting non-linear constitutive models, and it remains unclear if these parameters are robust to material variations and concentration changes, without a clear physical interpretation of the damping terms. Secondly, the temporal evolution of the filament thinning in the entangled regime progressively shifts away from the exponential response expected for a dilute solution \cite{entovEffectSpectrumRelaxation1997} towards a power-law relationship between the mid-plane radius and the time to singularity. This power-law relationship concomitantly results in transient extensional-thinning behavior in the extensional viscosity over a wide range of strain rates \cite{mckinleyViscoelastocapillaryThinningBreakup2005}. A phenomenological inelastic models have been applied by a recent study to explain this behavior \cite{dinicPowerLawsDominate2020}. However, the microstructural origin of this rate-thinning trend has not been considered, and a microscopically-orientated model that incorporates the evolution in the polymer conformation inside the thinning fluid element is lacking, inhibiting the extraction of accurate constitutive parameters from the filament thinning response. 

To address this limitation, in the present work, we calculate numerically the capillarity-driven thinning dynamics of entangled polymer solutions governed by two representative microstructural-based constitutive models derived from reptation theory. We show that the resulting filament thinning profiles exhibit a complex three-stage evolution dominated by tube reorientation, polymer chain stretching and ultimately the finite extensibility of a single polymer chain, respectively. The asymptotic solutions of the filament thinning profiles in each stage are calculated analytically. The predictions for the transient shear and extensional rheological response agree with the experimental observations for a number of entangled polymer solutions reported in previous studies without introducing additional fitting parameters. To illustrate the general trends, we present a dimensional prediction of the filament thinning profiles for a family of aqueous polyethylene oxide (PEO) solutions with varying concentrations in a transient extensional flow below, at, or above, the entangled threshold. The close similarity of the predicted filament thinning profiles with published experimental data provides additional physical insights into the process of the capillarity-driven thinning for an expanded variety of complex polymeric fluids.

\section{Numerical Model and Analysis}

\begin{figure}[!h]
    \centering
	\includegraphics[width=0.3\textwidth]{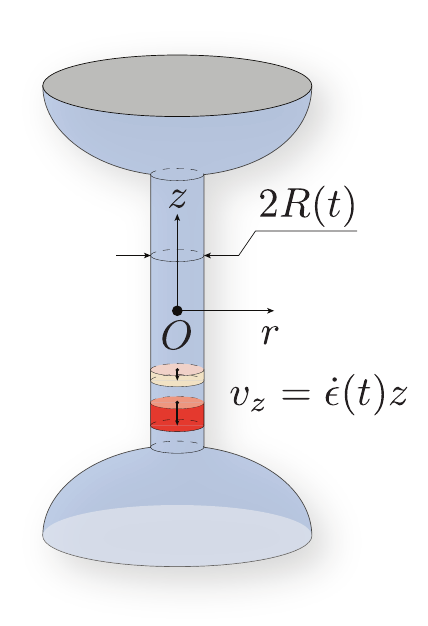}
    \caption{Geometrical setup of the capillarity-driven thinning process for a viscoelastic fluid. A thin cylindrical filament forms between two quasi-static hemispherical reservoirs and undergoes a self-thinning process driven by the capillary pressure and resisted by the viscoelasticity stress generated in the deforming thread. The extensional rheological parameters of the fluid can be extracted from measurements of the temporal evolution of the filament radius $R(t)$.}
    \label{fig:geometry}
\end{figure}

The geometrical configuration of the capillarity-driven thinning process is illustrated in Fig.~\ref{fig:geometry}. A long thin cylindrical liquid filament that exceeds the Rayleigh-Plateau stability limit for a given fluid volume and aspect ratio \cite{slobozhaninStabilityLiquidBridges1993} is generated by a rapid separation from the two coaxial discs. The temporal evolution of the filament radius $R(t)$ results from a balance between the capillary pressure and the viscoelastic stress arising from the fluid deformation.

In Fig.~\ref{fig:geometry}, the assumption of a slender cylindrical filament is valid when the filament thinning process is dominated by an elasto-capillary balance, and the stress components satisfy $\sigma_{zz}\gg\sigma_{rr}$  \cite{entovEffectSpectrumRelaxation1997,eggersPhysicsLiquidJets2008}. In the absence of inertial and gravitational effects, an explicit solution of $R(t)$ can be obtained by equating the material response to the filament thinning with the external driving stresses as \cite{mckinleyHowExtractNewtonian2000}
\begin{equation}
\eta_\mr{e}^+\dot{\epsilon}=N_1^+=\frac{F_\mr{a}(t)}{\pi R^2}-\Gamma K.
\label{eqn:stressbal}
\end{equation}
In this equation, the transient normal stress difference, $N_1^+=\sigma_{zz}^+-\sigma_{rr}^+$, describes the overall resistance to the filament thinning from the material response, which can be equivalently represented by the transient extensional viscosity $\eta_\mr{e}^+$. When the filament is close to cylindrical, the strain rate in a material element of size $\mr{d}V=\pi R^2\mr{d}z$ at any $z$ along the filament can be expressed \cite{mckinleyViscoelastocapillaryThinningBreakup2005} as
\begin{equation}
\dot{\epsilon}=-\dfrac{2\dot{R}(t)}{R(t)}.
\label{eqn:sr}
\end{equation}
The time-varying external axial force on the filament $F_\mr{a}(t)$ can be expressed from dimensional analysis in the form $F_\mr{a}(t)\equiv  2\pi X\Gamma R(t)$, where $X$ is a dimensionless geometric correction factor to account for the slenderness of filament shapes that are not perfectly cylindrical (\textit{i.e.}, $X=1$ for cylindrical shapes). The capillary pressure is derived from the surface tension $\Gamma$ and the mean curvature $K$ of the liquid filament, which can be expressed as
\begin{equation}
K=\dfrac{1}{R(1+R_{,z}^2)^{1/2}}+\dfrac{R_{,zz}}{(1+R_{,z}^2)^{3/2}}= \dfrac{1}{R},
\label{eqn:yl}
\end{equation}
where $R_{,z}\equiv\partial R/\partial z=0$ and $R_{,zz}\equiv \partial^2 R/\partial z^2=0$ due to the assumption of cylindrical filament shapes. 

From Eq.~\ref{eqn:stressbal}, the evolution of the filament radius under an elasto-capillary balance governed by the Hookean dumbbell model can be expressed \cite{entovEffectSpectrumRelaxation1997,
clasenHowDiluteAre2006,
wagnerAnalyticSolutionCapillary2015} as
\begin{equation}
R(t)=\left(\frac{GR_0^4}{2\Gamma}\right)^{1/3}\exp{\left(-\frac{t}{3\lambda}\right)},
\label{eqn:fene}
\end{equation}
where $G$ and $\lambda$ are the modulus and relaxation time characterizing the Hookean dumbbell constitutive equation, and $R_0$ is the filament radius at the onset of filament thinning. The Hookean dumbbell model has been applied extensively to describe the shear and transient extensional rheological responses for a number of dilute polymer solutions 
\cite{birdDynamicsPolymericLiquids1987,
roddCapillaryBreakupRheometry2005,
mckinleyViscoelastocapillaryThinningBreakup2005,
tirtaatmadja2006drop,
clasenHowDiluteAre2006}. In extensional flow at a constant strain rate, the Hookean dumbbell model shows an unbounded stress growth, which results in an exponential thinning in the radius $R(t)$ that is controlled by the relaxation time $\lambda$ as described by Eq.~\ref{eqn:fene}. This timescale is also consistent with the relaxation timescale in shear flow predicted by the same constitutive model. Finally, we note that substituting Eq.~\ref{eqn:fene} into the expression for $\dot{\epsilon}(t)$ given above results in a constant strain rate $\dot{\epsilon}=2/(3\lambda)$, or in a dimensionless form using the Weissenberg number as $\mr{Wi}\equiv \lambda\dot{\epsilon}=2/3$. 

\subsection{Tube models}
\subsubsection{Doi-Edwards-Marrucci-Grizzuti (DEMG) model}
\label{sec:demg}
Marrucci and Grizzuti \cite{marrucci1983free} modified the original Doi-Edwards (DE) model by introducing an additional evolution equation to describe the polymer chain stretch. As a result, in conjunction with the tube reorientation and chain reptation that are characterized by the disengagement time $\lambda_\mr{D}$, stretching of the flexible polymer chain introduces a second, independent relaxation process that becomes pivotal at higher strain rates \cite{geurts1988new}. This new relaxation process quantifies the stretching of the polymer chain along its primitive path length within an orientated tube segment and is characterized by the Rouse time of the polymer chain $\lambda_\mr{R}$. Pearson et al. \cite{pearsonTransientBehaviorEntangled1991} proposed a closed differential form of the constitutive equation for the DEMG model given by
\begin{subequations}
\begin{align}
\boldsymbol{S_{(1)}}&=-2(\boldsymbol{\nabla v}^\mr{T}:\boldsymbol{S})\boldsymbol{S}-\dfrac{1}{\Lambda^2 \lambda_\mr{D}}(\boldsymbol{S}-\dfrac{1}{3}\boldsymbol{I}), \\
\dot{\Lambda}&=\Lambda (\boldsymbol{\nabla v}^\mr{T}:\boldsymbol{S})-\dfrac{f(\Lambda)}{\lambda_\mr{R}}(\Lambda-1), \\
\boldsymbol{\sigma}&=3G_\mr{N} f(\Lambda)\Lambda^2 \boldsymbol{S}. 
\end{align}
\label{eqn:demg_model}%
\end{subequations}
In Eq.~\ref{eqn:demg_model}, $\boldsymbol{S}=\left<\boldsymbol{uu}\right>$ is the ensemble average of the tube orientation tensor, where $\boldsymbol{u}$ is the end-to-end vector for each tube segment. The scalar $\Lambda$ describes the overall strain imposed on the polymer chains within the tubes. The subscript ``$(1)$'' indicates the first-order upper-convected derivative \cite{birdDynamicsPolymericLiquids1987a}, and $\boldsymbol{\nabla v}$ is the velocity gradient tensor.
The finite extensibility factor for a single polymer chain $f(\Lambda)$ is given by the inverse Langevin function which can be approximated in a simple explicit form using Cohen's Pad\'e approximation \cite{cohen1991pade} as
\begin{equation}
f(\Lambda)=\frac{1-1/\Lambda_\mr{m}^2}{3-1/\Lambda_\mr{m}^2}\cdot\frac{3-\Lambda^2/\Lambda_\mr{m}^2}{1-\Lambda^2/\Lambda_\mr{m}^2},
\label{eqn:langevin}
\end{equation}
where $\Lambda_\mr{m}$ describes the maximum polymer extensibility which scales with the molecular weight as $\Lambda_\mr{m}\sim M^{1/2}$. From Eq.~\ref{eqn:demg_model}(a), it can be seen that the evolution of the orientation tensor $\boldsymbol{S}$ is a result of the interplay between the imposed convective flow and the tube relaxation characterized by the disengagement time $\lambda_\mr{D}$. By definition, the tube orientation tensor is constrained by $\Tr\boldsymbol{S}\equiv 1$ and $\boldsymbol{S}:\boldsymbol{S}\leq 1$. When $\boldsymbol{S}=\boldsymbol{I}/3$, all tube segments are, on average, randomly oriented \cite{marrucci1983free}.

To calculate the capillarity-driven thinning dynamics of a thin fluid filament described by the DEMG model, we substitute Eq.~\ref{eqn:demg_model} into Eq.~\ref{eqn:stressbal}. For mathematical simplicity, we non-dimensionalize the expression according to $\h{R}\equiv R/R_0$ and $\h{t}\equiv t/\lambda_\mr{R}$, where $R_0=R(0)$ is the filament radius at $t=0$. 
The dimensionless stress balance equation can be expressed as
\begin{equation}
\dfrac{1}{\h{R}}=3 \mr{Ec}_0 f(\Lambda)\Lambda^2\Delta S,
\label{eqn:demg_bal}
\end{equation}
where $\Delta S\equiv S_{zz}-S_{zz}$ is the magnitude of the tube reorientation calculated from the difference of the $zz$- and $rr$-components of the reorientation tensor, and $\mr{Ec}_0\equiv G_\mr{N} R_0/\Gamma$ is the intrinsic elasto-capillary number of the fluid, where $G_\mr{N}$ is the viscoelastic plateau modulus of the entangled fluid system. Because of the axisymmetric shape of the filament and the irrotational nature of the extensional flow, the $rr$- and $\theta\theta$-components of the reorientation tensor are identical in magnitude. Therefore, the trace of the reorientation tensor satisfies $\Tr\boldsymbol{S}=S_{zz}+2S_{rr}= 1$, and the temporal evolution of $\Delta S$ can be rewritten as
\begin{equation}
\Delta S_{,\h{t}}=\mr{Wi}(\Delta S + 1) -2\mr{Wi}\Delta S^2-\dfrac{1}{\Lambda^2 p(Z)}\Delta S.
\label{eqn:demg_w}
\end{equation}
Here, the Weissenberg number quantifying the strength of the flow is given by $\mr{Wi}\equiv \lambda_\mr{R}\dot{\epsilon}=-2\h{R}_{,\h{t}}/\h{R}$ according to Eq.~\ref{eqn:sr}. The ratio of timescales in Eq.~\ref{eqn:demg_w} as $p(Z)=\lambda_\mr{D}/\lambda_\mr{R}$ can be expressed as a function of the number of entanglements per polymer chain $Z$ as calculated from Eq.~\ref{eqn:Z}.

\subsubsection{Rolie-Poly (RP) model}
In addition to the finite rates of polymer chain stretching and retraction within the tube that are incorporated in the DEMG model, convective constraint release (CCR) has been recognized more recently as another critical non-reptative mechanism that alters the non-linear rheology of entangled polymer systems in both steady and transient flows \cite{marrucciDynamicsEntanglementsNonlinear1996,
bhattacharjeeExtensionalRheometryEntangled2002,
grahamMicroscopicTheoryLinear2003,
dealyMolecularStructureRheology2006}. CCR is manifested at $\dot{\epsilon}>1/\lambda_\mr{D}$, when polymer disengagement from the physically entangled network becomes faster than reptation due to the surrounding polymer molecules being convected away from the chain of interest by the strong flow. To obtain a more comprehensive understanding of how CCR potentially affects the dynamics of capillarity-driven thinning, we adopt the more sophisticated Rolie-Poly (RP) model proposed by Likhtman and Graham \cite{likhtmanSimpleConstitutiveEquation2003}, and compare the resulting temporal evolution of the filament thinning profiles with those obtained from the DEMG model. A specific form of the constitutive equation similar to Eq.~\ref{eqn:demg_model} is taken that separates the tube reorientation and the polymer chain stretch \cite{dealyMolecularStructureRheology2006} according to
\begin{subequations}
\begin{align}
\boldsymbol{S_{(1)}}&=-2(\boldsymbol{\nabla v}^\mr{T}:\boldsymbol{S})\boldsymbol{S}-\dfrac{1}{\Lambda^2}\left[\left(\dfrac{1}{\lambda_\mr{D}}+2\beta f(\Lambda)\dfrac{1-1/\Lambda}{\lambda_\mr{R}}\Lambda^{\delta}\right)(\boldsymbol{S}-\dfrac{1}{3}\boldsymbol{I})\right], \\
\dot{\Lambda}&=\Lambda (\boldsymbol{\nabla v}^\mr{T}:\boldsymbol{S})-\dfrac{f(\Lambda)}{\lambda_\mr{R}}(\Lambda-1)-\left(\dfrac{1}{\lambda_\mr{D}}+2\beta f(\Lambda)\dfrac{1-1/\Lambda}{\lambda_\mr{R}}\Lambda^{\delta}\right)\dfrac{\Lambda^2-1}{2\Lambda}, \\
\boldsymbol{\sigma}&=3G_\mr{N} f(\Lambda) \Lambda^2 \boldsymbol{S},
\end{align}
\label{eqn:roliepoly}%
\end{subequations}
where the notations of $\boldsymbol{S}$, $\Lambda$ and $\boldsymbol{\nabla v}$ are identical to those in Eq.~\ref{eqn:demg_model}, and the Pad\'e approximation for the inverse Langevin function $f=f(\Lambda)$ is evaluated from Eq.~\ref{eqn:langevin}. The dimensionless coefficients $\beta$ and $\delta$ are two model parameters that quantify the magnitude of the CCR effect and the contribution of the polymer chain stretch to the CCR effect, respectively. Previous studies have suggested that $\beta=1$ and $\delta=-0.5$ produce the best-fitting results with the full kinetic model over a wide range of shear rates \cite{marrucciDynamicsEntanglementsNonlinear1996, likhtmanSimpleConstitutiveEquation2003}, and we thus adopt these values in our calculation and asymptotic analysis. 

Following the numerical procedure for the DEMG model, the differential equation for the temporal evolution of the axial averaged tube orientation $\Delta 
S\equiv S_{zz}-S_{rr}$ can be rewritten in a similar form as Eq.~\ref{eqn:demg_w} with the addition of the CCR term as 
\begin{equation}
\Delta S_{,\h{t}}=\mr{Wi}(\Delta S + 1) -2\mr{Wi}\Delta S^2-\dfrac{1}{\Lambda^2}\left[\dfrac{1}{p(Z)}+2\beta f(\Lambda)(1-1/\Lambda)\Lambda^{\delta}\right]\Delta S,
\label{eqn:rp_w}
\end{equation}
where the time-varying Weissenberg number is identically defined as in Eq.~\ref{eqn:demg_w}, and the stress balance given by Eq.~\ref{eqn:demg_bal} remains unmodified. 
\subsection{Capillarity-driven thinning dynamics}
\label{sec:analytiacal}
\begin{figure*}[!]
    \centering
	\includegraphics[width=\textwidth]{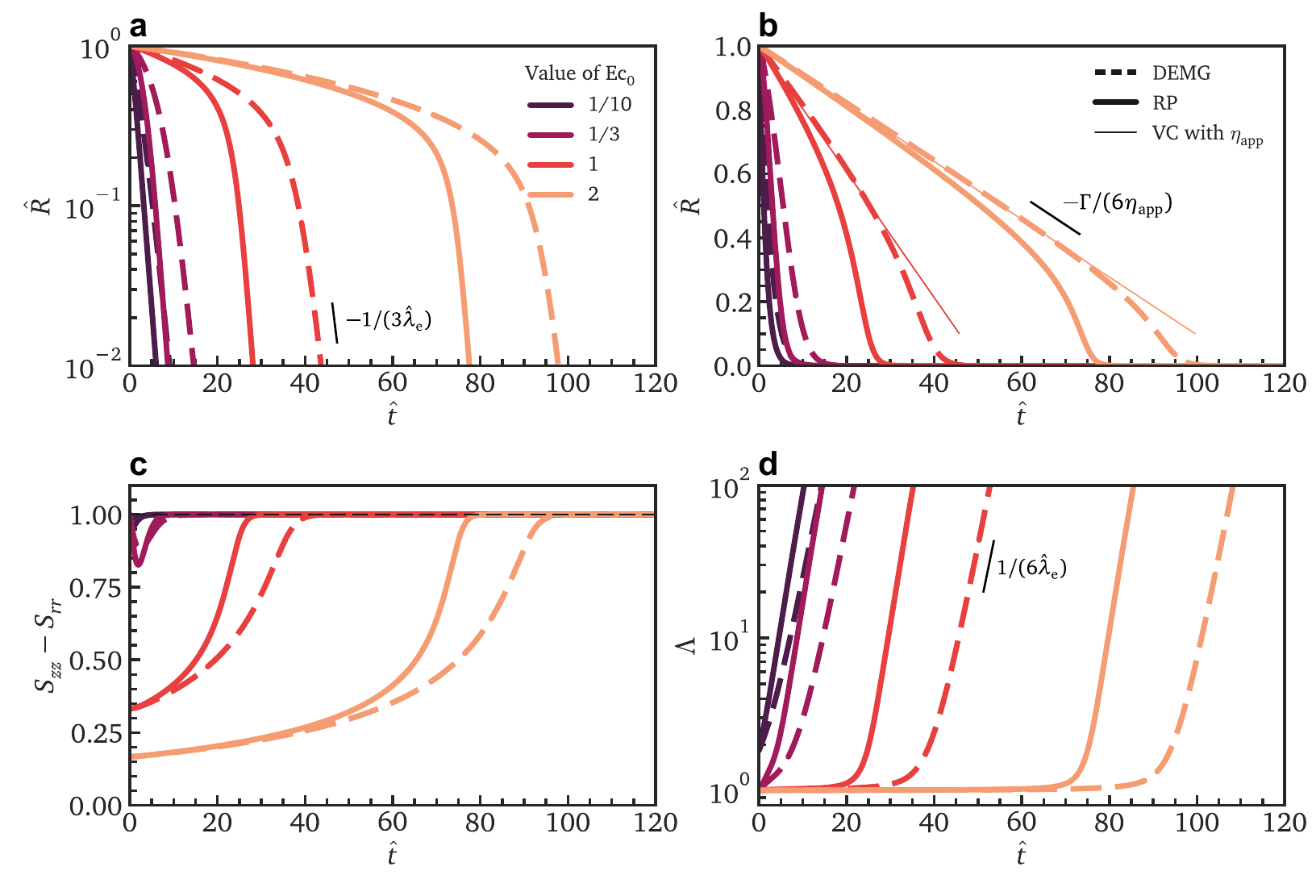}
    \caption{Capillarity-driven thinning dynamics described by the two selected models (dashed line: DEMG model; solid line: Rolie-Poly model) with varying elasto-capillary numbers $\mr{Ec}_0=1/10$, $1/3$, $1$ and $2$ at a fixed number of entanglements per polymer chain, $Z=10$ ($\lambda_\mr{D}/\lambda_\mr{R}=8.97$). Here infinite extensibility of the polymer chain ($\Lambda_\mr{m}\rightarrow\infty$) is chosen for numerical simplicity. (a) The temporal evolution of the dimensionless filament radius $\h{R}\equiv R/R_0$ against $\h{t}\equiv t/\lambda_\mr{R}$ on a logarithmic scale predicted by both selected models. When $\mr{Ec}_0>1/3$, the filament thinning profiles exhibit two distinct stages. All the curves show an exponentially-thinning trend when the filament radius is sufficiently small. (b) The temporal evolution of the dimensionless radius ($\h{R}\equiv R/R_0$) predicted by both selected models on a linear scale. The thinning profile expected for a Newtonian fluid, using the apparent viscosity from Eq.~\ref{eqn:eta0} is plotted for $\mr{Ec}_0=1$ and $\mr{Ec}_0=2$. (c) The temporal evolution of the tube orientation $\Delta S\equiv S_{zz}-S_{rr}$ predicted by both selected models. (d) The temporal evolution of polymer chain stretch $\Lambda$ showing an exponential growth at long times, which is delayed when $\mr{Ec}_0>1/3$.  
}
    \label{fig:early_thinning}
\end{figure*}

Numerical calculations of the capillarity-driven thinning dynamics for the two selected constitutive models have been performed for a range of representative material parameters. The initial conditions for the reorientation tensor and the chain stretch following the initial axial step strain that form that liquid bridge are shared by both models. The reasons for these choices will be explained in detail in the rest of this manuscript, and they are specified according to the magnitude of the elasto-capillary number ($\mr{Ec}_0$) as
\begin{subequations}
\begin{align*}
&\mr{Ec}_0\geq \dfrac{1}{3}: &&S_{rr}(0)=\dfrac{1}{3}\left(1-\dfrac{1}{3\mr{Ec}_0}\right), &&S_{zz}(0)=\dfrac{1}{3}\left(1+\dfrac{2}{3\mr{Ec}_0}\right), &&\Lambda(0)=1, \\
&\mr{Ec}_0< \dfrac{1}{3}: &&S_{rr}(0)=0, &&S_{zz}(0)=1, &&f[\Lambda(0)]\Lambda(0)^2=\dfrac{1}{3\mr{Ec}_0}.
\end{align*}
\label{eqn:init}%
\end{subequations}
In Fig.~\ref{fig:early_thinning}, the temporal evolution of the filament radius, the tube reorientation and the chain stretch for both models (dashed line: DEMG model; solid line: RP model) are presented. Here the elasto-capillary number $\mr{Ec}_0$ varies from $\mr{Ec}_0=\numrange{0.1}{2}$, and the number of entanglements is fixed at $Z=10$. Here we initially set $\Lambda_\mr{m}\rightarrow\infty$ because the finite extensibility of the polymer chain only affects the filament thinning behavior very close to the filament breakup and is thus not depicted in this figure. The overall trends observed in the filament radius described by the two models are similar, except that the RP model predicts a noticeably faster filament thinning due to the additional CCR effect. By inspecting the temporal evolution of the filament thinning profiles at $\mr{Ec}_0=2$ (dark purple lines), we find two distinct filament thinning regimes. At an early time, the filament radius decays slowly. In this stage, the magnitude of tube reorientation $\Delta S\equiv S_{zz}-S_{rr}$ increases progressively to unity, indicating full alignment in the extensional direction; the polymer molecules however remain initially unstretched. As the magnitude of the tube reorientation saturates to unity, the filament radius starts to decay at a much faster rate in an exponential manner. The filament kinematics in this fast-thinning stage can be attributed to the rapid polymer chain stretch. When $\mr{Ec}_0=0.1<1/3$ (orange lines), the normal stress generated by the tube orientation does not suffice to balance the driving capillary pressure (\textit{i.e.}, Eq.~\ref{eqn:demg_bal} cannot be satisfied as written), and a nontrivial polymer stretch at $t=0$ is necessary to produce a larger normal stress difference to balance the driving capillary pressure, as specified from Eq.~\ref{eqn:init}. Consequently, only the second exponential-thinning regime is manifested.

\begin{figure*}[!]
    \centering
	\includegraphics[width=\textwidth]{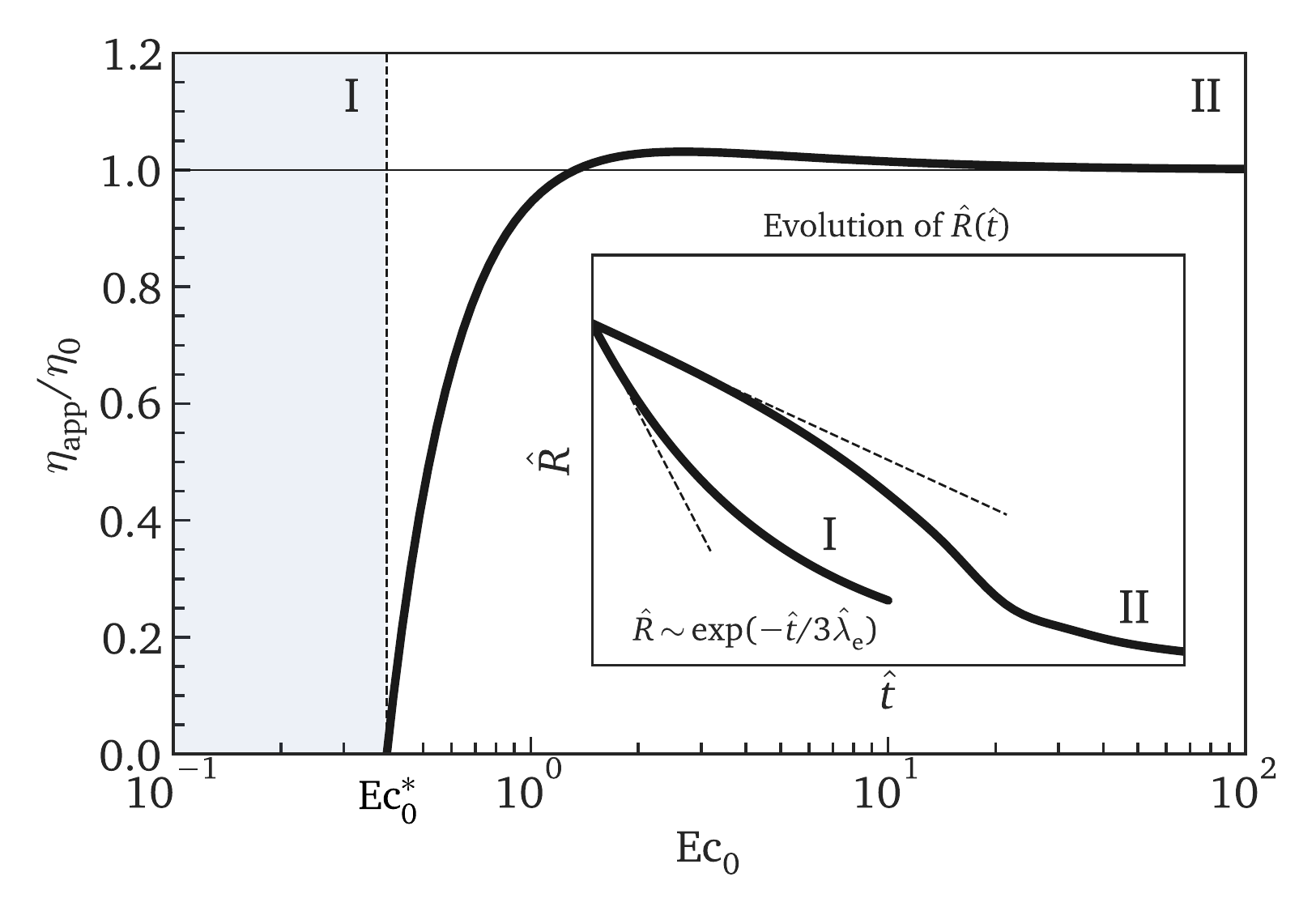}
    \caption{Apparent shear viscosity $\eta_\mr{app}(\mr{Ec}_0)$ obtained from the filament thinning profiles in the early thinning regime. A meaningful viscosity reading (which must be positive) is obtained for $\mr{Ec}_0>\mr{Ec}_0^*$ (indicated by the thin dashed line). As $\mr{Ec}_0$ increases, the apparent shear viscosity measured by the filament thinning process approaches the zero-shear viscosity $\eta_0=G_\mr{N}\lambda_\mr{D}$ (thin solid line). Inset: Schematic of the evolution of filament radius for $\mr{Ec}_0<\mr{Ec}_0^*$ (I) and $\mr{Ec}_0\geq \mr{Ec}_0^*$ (II). The dashed lines show reference lines where $R(t)$ decreases linearly with $t$.}
    \label{fig:eta_0}
\end{figure*}

When plotted on a linear scale as shown in Fig.~\ref{fig:early_thinning}(b), the temporal evolution of the filament radius exhibits a linear decrease for $\h{R}\gtrsim 0.4$ when $\mr{Ec}_0$ is sufficiently large. Notably, the filament kinematics in the early thinning regime are similar to those predicted from a Newtonian fluid \cite{mckinleyHowExtractNewtonian2000} with a thinning rate given by $\dot{R}\sim-\Gamma/\eta_\mr{app}$. In practice, a measure of the shear viscosity at low strain rates can be obtained from this linear trend assuming the filament to be of cylindrical shape. This apparent shear viscosity can be analytically calculated by imposing $\Lambda=1$ in the constitutive equations (Eq.~\ref{eqn:demg_model} and Eq.~\ref{eqn:roliepoly}) as
\begin{equation}
\mr{Tr}_{\mr{app}}\equiv \dfrac{\eta_\mr{app}}{G_\mr{N}\lambda_\mr{D}}=1+\dfrac{1}{6\mr{Ec}_0} - \dfrac{2}{9\mr{Ec}_0^2}.
\label{eqn:eta0}
\end{equation}
Eq.~\ref{eqn:eta0} is further plotted against $\mr{Ec}_0$ in Fig.~\ref{fig:eta_0}. From this figure, we notice a threshold of $\mr{Ec}_0>\mr{Ec}_0^*=8/(3+\sqrt{297})\approx 0.395$ when a valid apparent viscosity (positive) is obtained. This condition is substantially identical to the piecewise initial conditions in Eq.~\ref{eqn:init}, where the tube reorientation dominates the filament thinning dynamics in the early thinning regime for $\mr{Ec}_0>1/3$. As $\mr{Ec}_0$ increases, this apparent shear viscosity $\eta_\mr{app}$ approaches the zero-shear viscosity given by $\eta_0=G_\mr{N}\lambda_\mr{D}$. 

\begin{figure}[!]
    \centering
    \includegraphics[width=\textwidth]{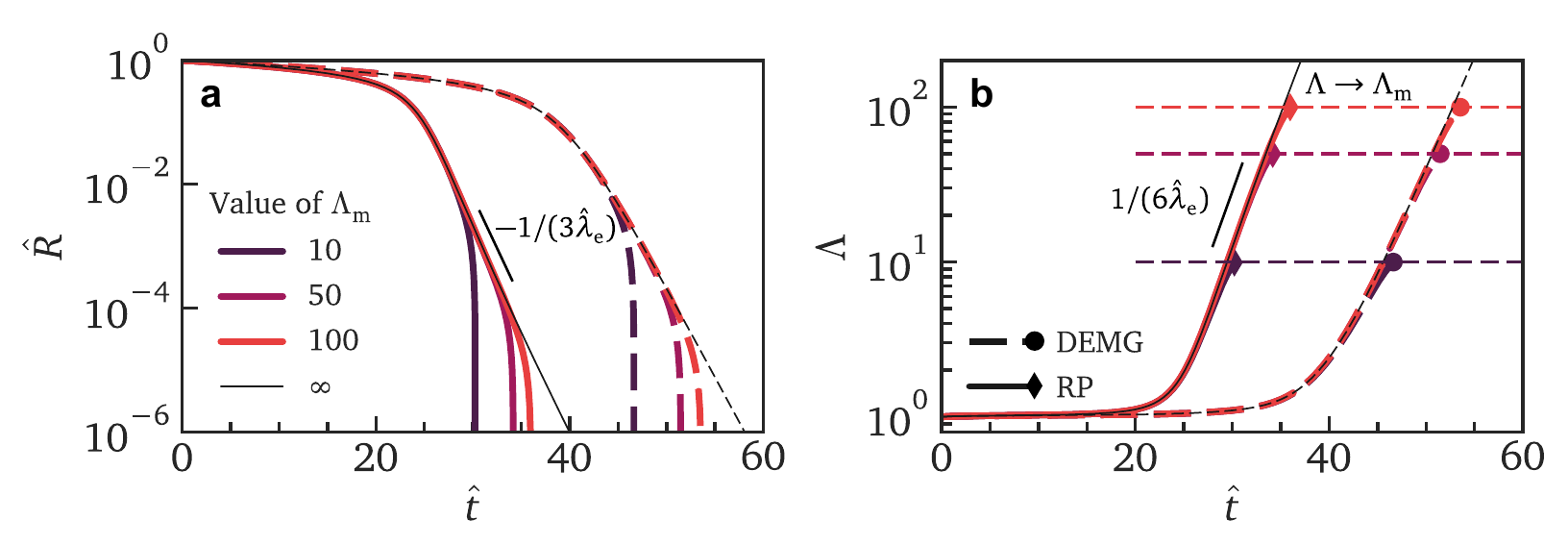}
    \caption{Capillarity-driven thinning dynamics described by the two selected models (dashed line: DEMG model; solid line: Rolie-Poly model) with varying finite extensibility factors $\Lambda=\numlist{10;50;100}$ and $\Lambda_\mr{m}\rightarrow\infty$ at a fixed elasto-capillary number, $\mr{Ec}_0=1$ and a fixed number of entanglements per polymer chain, $Z=10$. (a) The temporal evolution of the dimensionless filament radius $\h{R}\equiv R/R_0$ on a logarithmic scale at intermediate times. A consistent exponential thinning trend is manifested for both models at intermediate times; at longer times, the filament radius deviates substantially from the exponential-thinning trend and rapidly approaches singularity. An apparent extensional relaxation time $\h{\lambda}_\mr{e}$ (nondimensionalized by $\lambda_\mr{R}$) can be obtained from fitting the exponentially-decaying region. A practical optical resolution limit is indicated at $\h{R}=0.001$. (b) The temporal evolution of the polymer chain stretch $\Lambda$ on a logarithmic scale at intermediate times. Initially the stretch increases exponentially; close to the point of filament breakup, the polymer chain stretch $\Lambda$ approaches the specified maximum stretch of $\Lambda_\mr{m}$ (thin horizontal dashed lines). 
  }
    \label{fig:intermediate_thinning}
\end{figure}

The exponential-thinning regime observed from the filament thinning profiles of both models at intermediate time, as shown in Fig.~\ref{fig:early_thinning}, is reminiscent of the filament thinning dynamics under the elasto-capillary balance predicted by the Hookean dumbbell model for dilute polymer solutions. Nevertheless, it remains unclear if the polymer relaxation process in the exponential-thinning trend presented in this figure is consistent with that expected for the dilute polymer systems. To obtain an accurate measure of the extensional rheological properties from the capillarity-driven thinning dynamics, we revisit the filament thinning profiles of both selected models with varying finite extensibilities $\Lambda_\mr{m}$, a fixed elasto-capillary number $\mr{Ec}_0=1$ and a fixed number of entanglements per polymer chain, $Z=10$. As shown in Fig.~\ref{fig:intermediate_thinning}, a consistent exponential-thinning trend can be observed at intermediate time for the both models with an identical thinning rate. From previous discussions, this exponential-thinning trend in the filament radius is dominated by the exponentially increasing polymer chain stretch $\Lambda$ induced by a strong extensional flow. As $\Lambda$ approaches the specified finite maximum chain stretch $\Lambda_\mr{m}$, the temporal evolution of the filament radius substantially deviates from the exponential thinning trend and the thread breaks in finite time. 

\begin{figure}[!]
    \centering
    \includegraphics[width=\textwidth]{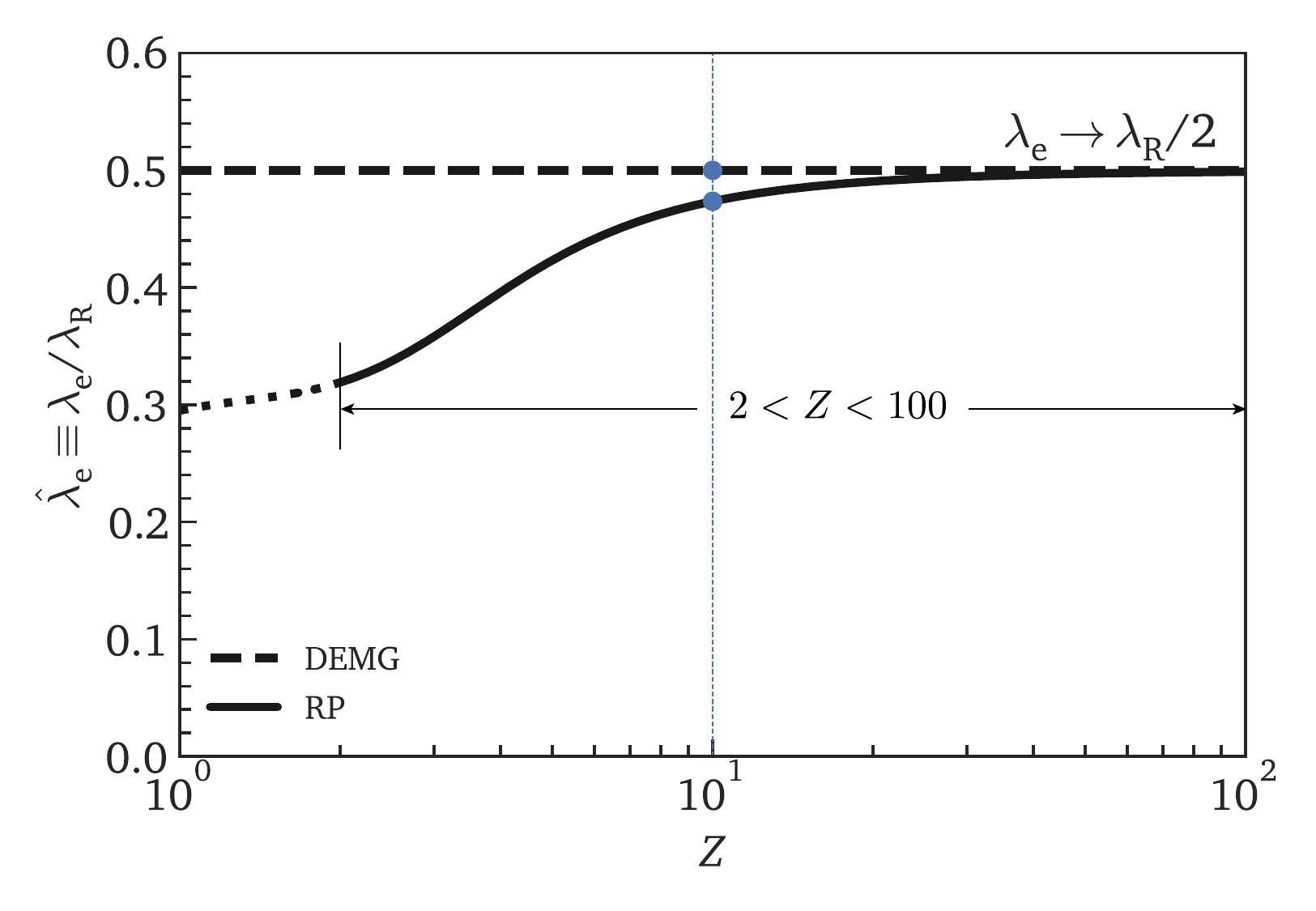}
    \caption{The dimensionless apparent extensional relaxation time $\h{\lambda}_\mr{e}=\lambda_\mr{e}/\lambda_\mr{R}$ extracted from the exponential-thinning regime of the filament thinning dynamics for both selected models. For the DEMG model (dashed line), exponentially fitting always gives $\lambda_\mr{e}=\lambda_\mr{R}/2$. For the Rolie-Poly model (solid line), a monotonic trend is observed, approaching the DEMG model as $Z\rightarrow\infty$. The result for $Z<2$ is shown as a dashed line due to the dubious accuracy of the Taylor series expansion (Eq.~\ref{eqn:Z}) in this limit \cite{likhtmanQuantitativeTheoryLinear2002}. The blue marker shows the result for the Rolie-Poly model at $Z=10$ used in Fig.~\ref{fig:early_thinning}, \ref{fig:intermediate_thinning} and \ref{fig:final_thinning}.}
    \label{fig:ext_relaxation}
\end{figure}

While the capillarity-driven thinning dynamics predicted by these tube models for entangled polymer solutions are distinct from those predicted by the Hookean dumbbell models for dilute polymer solutions, it is common in practice to extract an extensional relaxation time from the exponential-thinning regime. To mitigate the ambiguity in the interpretation of the measurements from the application of different models, we define an \textit{apparent} extensional relaxation time $\h{\lambda}_\mr{e}$ (nondimensionalized by $\lambda_\mr{R}$) that is obtained from fitting the data using the Hookean dumbbell model, such that in the exponential-thinning regime, $\h{R}(\h{t})\sim \exp[-\h{t}/(3\h{\lambda}_\mr{e})]$. For each selected models, an analytical solution of this exponential-thinning process can be obtained by solving Eq.~\ref{eqn:demg_model}, Eq.~\ref{eqn:demg_bal} and Eq.~\ref{eqn:roliepoly} in the limit of $\mr{Ec}_0\rightarrow 0$ (negligible tube orientation) and $\Lambda_\mr{m}\rightarrow \infty$ (infinite extensibility for polymer chains). The \textit{apparent} extensional relaxation times characterizing the exponential thinning process in each model are thus found to be
\begin{subequations}
\begin{align}
\h{\lambda}_\mr{e,DEMG}&\equiv \dfrac{\lambda_\mr{e,DEMG}}{\lambda_\mr{R}}=\dfrac{1}{2}, \\
\h{\lambda}_\mr{e,RP}&\equiv \dfrac{\lambda_\mr{e,RP}}{\lambda_\mr{R}}=\dfrac{1}{2+1/p(Z)},
\label{eqn:rel_time}
\end{align}
\end{subequations}
where $p(Z)=\lambda_\mr{D}/\lambda_\mr{R}$ is the ratio of the disengagement time and the Rouse time expressed in Eq.~\ref{eqn:Z}. These expressions are plotted in Fig.~\ref{fig:ext_relaxation}. It is evident that the apparent extensional relaxation times $\lambda_\mr{e}$ measured from the filament thinning profiles are primarily determined by the magnitude of the Rouse time $\lambda_\mr{R}$ in the entangled fluid. As $Z\rightarrow\infty$, both expressions for $\lambda_\mr{e}$ approach an identical limit of $\lambda_\mr{R}/2$. The apparent extensional relaxation time from the RP model includes a coupling term that depends on the disengagement time $\lambda_\mr{D}$ (through $p(Z)$). This term arises from CCR but only becomes comparable in magnitude with the Rouse time contribution when $Z$ is relatively small. In this limit, polymer chain relaxation is further promoted by the convected displacement of surrounding polymer molecules in the transient extensional flow. 

\begin{figure}[!]
    \centering
    \includegraphics[width=\textwidth]{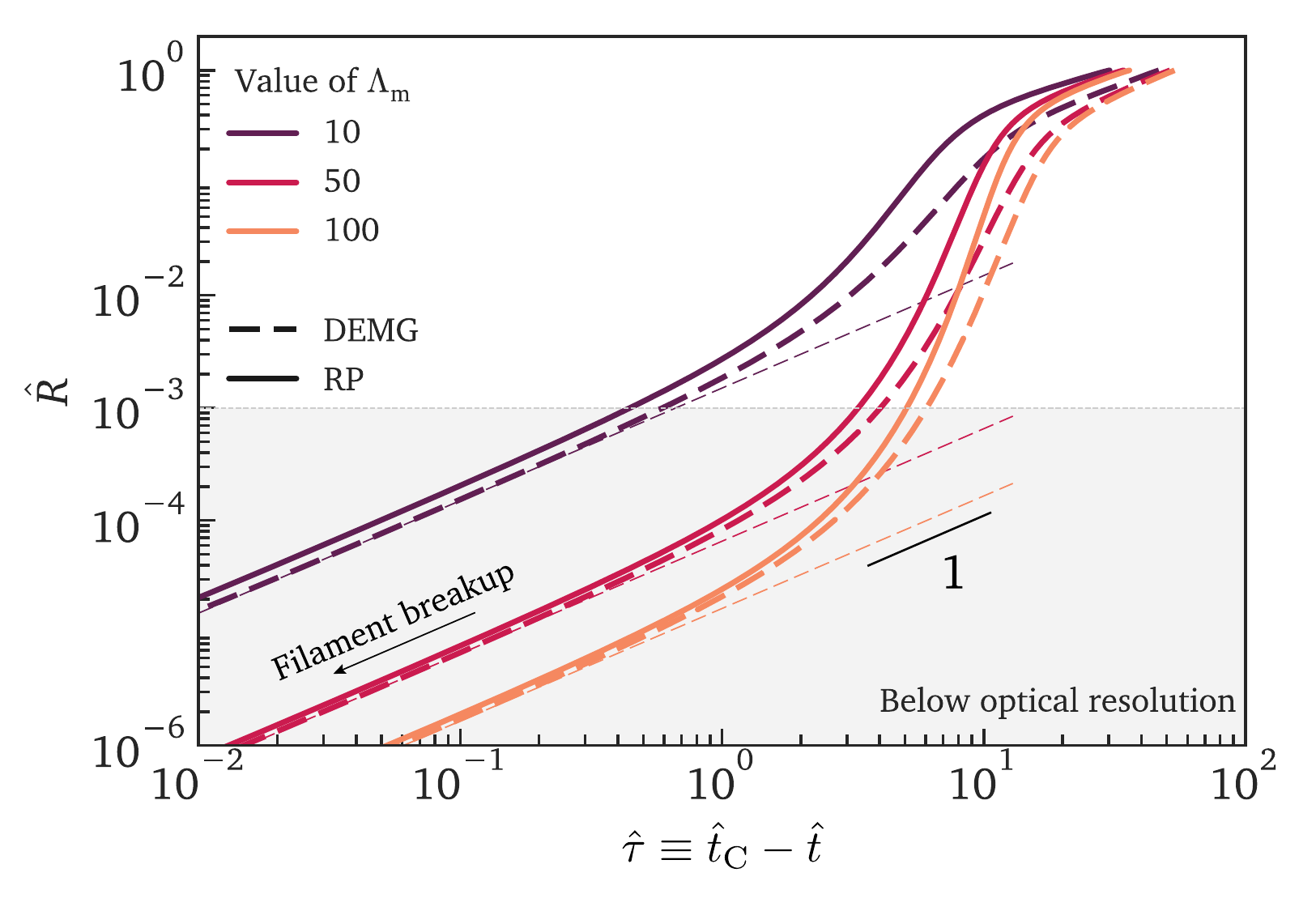}
    \caption{Evolution of the filament radius close to breakup with varying finite extensibility factors $\Lambda_\mr{m}=\numlist{10;50;100}$ at a fixed elasto-capillary number $\mr{Ec}_0=1$ and fixed number of entanglements per polymer chain $Z=10$. The filament radius is plotted against $\h{\tau}\equiv \h{t}_\mr{C}-\h{t}$, where $\h{t}_\mr{C}$ is critical time of breakup (nondimensionalized by $\lambda_\mr{R}$). Linear decrease of the filament radius is observed close the breakup, consistent with the filament thinning response of a Newtonian fluid. As a result, a terminal extensional viscosity $\eta_\mr{e,\infty}$ can be obtained analytically close to singularity. Asymptotic solutions of the filament radius are plotted as thin dashed lines. A practical optical limit of $\h{R}=0.001$ is also indicated.}
    \label{fig:final_thinning}
\end{figure}

Finally, we numerically calculate the filament thinning profiles close to the final breakup event, where the polymer chain stretch $\Lambda$ approaches the maximum extensibility $\Lambda_\mr{m}$ and the temporal evolution of the filament radius deviates from the exponential-thinning trend. For clarity in the subsequent analysis, we transform the time axis by plotting the results against $\h{\tau}\equiv\h{t}_\mr{C}-\h{t}$, where $\h{t}_\mr{C}$ is the time of filament breakup (nondimensionalized by $\lambda_\mr{R}$) which can be numerically obtained from extrapolation of the filament radius to $\h{R}=0$. As shown in Fig.~\ref{fig:final_thinning}, the temporal evolution of $\h{R}(\h{t})$ with the same parameter values as in Fig.~\ref{fig:intermediate_thinning} is plotted for the two selected models. As the filament approaches singularity ($\h{t}\rightarrow \h{t}_\mr{C}$, or $\h{\tau}\rightarrow 0$), finite polymer extensibility progressively dominates the filament thinning dynamics. Consequently, the filament radius undergoes a linear decay, and the asymptotic predictions of the two models at identical values of finite extensibility $\Lambda_\mr{m}$ overlap with each other.

The linear decay of the filament radius close to the filament breakup has also been observed for the FENE-P model \cite{wagnerAnalyticSolutionCapillary2015}. This asymptotic linear variation is consistent with the capillarity-driven thinning dynamics of a highly anisotropic Newtonian fluid \cite{mckinleyHowExtractNewtonian2000}. Consequently, we can extract the terminal extensional viscosity $\eta_\mr{e,\infty}$ predicted by the models in this linearly-thinning regime by rewriting Eq.~\ref{eqn:stressbal} and Eq.~\ref{eqn:demg_bal} in form of Taylor expansion regarding to $1/\Lambda$ (where $\Lambda<\Lambda_\mr{m}$). Consequently, an analytical expression for the terminal extensional viscosity can be obtained as
\begin{equation}
\eta_\mr{e,\infty}=\dfrac{3G_\mr{N} \lambda_\mr{R} \Lambda_\mr{m}^2}{1-1/\Lambda_\mr{m}}.
\label{eqn:terminal_vis}
\end{equation}
The asymptotic solutions for the filament radius using $\eta_\mr{e,\infty}$ with varying finite extensibilities $\Lambda_\mr{m}$ are shown in Fig.~\ref{fig:final_thinning} as thin dashed lines, and they are consistent with the full numerical calculations of the filament thinning dynamics close to breakup. 

\section{Discussion}
\subsection{Ratio of the apparent extensional and shear relaxation times}
\label{sec:comp}
In practice, the shear relaxation time $\lambda_\mr{s}$ is routinely obtained from the small amplitude oscillatory shear (SAOS) as $1/\omega_\mr{c}$, where $\omega_\mr{c}$ is the characteristic angular frequency at which the storage and loss moduli cross. For model dilute, monodisperse polymer solutions that can be well described by the Hookean or FENE dumbbell model, there is a single characteristic relaxation time that characterizes the polymer chain dynamics regardless of the flow type (shear or extension). However, a number of previous studies of the capillarity-driven thinning dynamics observed in different entangled polymer solutions have revealed distinctly different measures of the relaxation time in shear and extensional flow \cite{arnoldsCapillaryBreakupExtensional2010,
sachsenheimerExperimentalStudyCapillary2014,
sachsenheimer2014elongational,
dinicPowerLawsDominate2020}. A simple dumbbell model cannot readily reconcile these differences in the reported relaxation times from a physical perspective. 

In this section, we establish a theoretical framework using the tube models that allows us to clearly interpret the difference in shear and extensional relaxation times observed in entangled polymer solutions. We have shown in Section~\ref{sec:analytiacal} that the temporal evolution of the filament radius resulted from capillarity-driven thinning features an exponential-thinning trend at intermediate time. In this stage, chain extension and the Rouse time $\lambda_\mr{R}$ dominate the filament thinning process and controls the magnitude of the \textit{apparent} extensional relaxation time. In contrary, the shear relaxation time can be analytically calculated by imposing a small-amplitude oscillatory shear strain to the DEMG and the RP models. If only the leading-order terms are retained, both tube models reduce to the Maxwell model with a longest relaxation time $\lambda_\mr{D}$ and modulus $G=3G_\mr{N}$. This result is not surprising, because at small strains, the rheology is dominated by the processes of tube deformation and reorientation, which is consistent with the relaxation mode described by the disengagement time $\lambda_\mr{D}$. To summarize, we compare the expected apparent shear ($\lambda_\mr{s}$) and extensional ($\lambda_\mr{e}$) relaxation times of the Hookean dumbbell model and the two selected tube models in Table~\ref{table:SAOS}. Without introducing additional fitting parameters, we can quantify the ratio of the two relaxation times using the constitutive parameters of the tube models.

\begin{table*}[!ht]
\centering
\caption{Dynamic moduli $G'$ and $G''$, the crossover angular frequency $\omega_\mr{c}$ where $G'=G''$, and the \textit{apparent} shear and extensional relaxation times $\lambda_\mr{s}=1/\omega_\mr{c}$ and $\lambda_\mr{e}$ for the Hookean dumbbell model, the DEMG model, and the Rolie-Poly model. The apparent extensional relaxation time in the limit of $Z\rightarrow \infty$ is also shown.}

\begin{tabular*}{\textwidth}{p{3.5cm}p{2cm}p{2cm}p{0.8cm}p{0.8cm}p{2cm}p{3cm}}
\toprule
Model & $G'(\omega)$ & $G''(\omega)$ & $\omega_\mr{c}$ & $\lambda_\mr{s}$ & $\lambda_\mr{e}$ & $\lambda_\mr{e}(Z\rightarrow\infty)$\\
\midrule
Hookean dumbbell & $\dfrac{G\lambda^2\omega^2}{1
+(\lambda\omega)^2}$ & $\dfrac{G\lambda\omega}{1
+(\lambda\omega)^2}$ & $\dfrac{1}{\lambda}$ & $\lambda$ & $\lambda$ & $\lambda$\\\addlinespace[1em]

DEMG & $\dfrac{G_\mr{N}\lambda_\mr{D}^2 \omega^2}{1
+(\lambda_\mr{D}\omega)^2}$ & $\dfrac{G_\mr{N}\lambda_\mr{D}\omega}{1
+(\lambda_\mr{D}\omega)^2}$ & $\dfrac{1}{\lambda_\mr{D}}$ & $\lambda_\mr{D}$ & $\dfrac{1}{2}\lambda_\mr{R}$ & $\dfrac{1}{2}\lambda_\mr{R}$\\\addlinespace[1em]

Rolie-Poly & $\dfrac{G_\mr{N}\lambda_\mr{D}^2\omega^2}{1
+(\lambda_\mr{D}\omega)^2}$ & $\dfrac{G_\mr{N}\lambda_\mr{D}\omega}{1
+(\lambda_\mr{D}\omega)^2}$ & $\dfrac{1}{\lambda_\mr{D}}$ & $\lambda_\mr{D}$ & $\dfrac{\lambda_\mr{R}}{2+1/p(Z)}$ & $\dfrac{1}{2}\lambda_\mr{R}$\\\addlinespace[0.5em]

\bottomrule
\end{tabular*}

\label{table:SAOS}
\end{table*}


For entangled polymer solutions, the molecular weight between entanglements $M_\mr{e}(c)$ is, of course, larger than reported for polymer melts due to dilution of the polymer chains. To incorporate the dependence on polymer concentration in our expression for the relaxation time ratio, we refer to Pearson et al. \cite{pearsonFlowInducedBirefringence1989} and Ravindranath et al. \cite{ravindranathBandingSimpleSteady2008} and calculate the effective molecular weight between entanglements in the entangled solutions as
\begin{equation}
M_\mr{e}(c)=M_\mr{e,0}c^{\frac{1}{1-3\nu}},
\label{eqn:me_sol}
\end{equation}
where $c$ is the mass fraction of the polymers in solution, and $M_\mr{e,0}$ is the molecular weight between entanglements in the polymer melts ($c=1$). The excluded volume parameter $\nu=0.5$ for $\theta$-solvents, and $\nu=0.6$ for good solvents. From Eq.~\ref{eqn:Z}, the number of entanglements per chain in the entangled polymer solutions can be there expressed as
\begin{equation}
Z_\mr{sol}(c)=\dfrac{M}{M_\mr{e,0}}c^{\frac{1}{3\nu-1}}\equiv Z_0 c^{\frac{1}{3\nu-1}}.
\label{eqn:Zphi}
\end{equation}
From Eq.~\ref{eqn:Zphi}, we can thus evaluate the evolution in relaxation time ratio $\lambda_\mr{e}/\lambda_\mr{s}$ as a function of the number of entanglements per chain $Z_\mr{sol}(c)$ as
\begin{equation}
\dfrac{\lambda_\mr{e}}{\lambda_\mr{s}}=\dfrac{\h{\lambda}_\mr{e}[Z_\mr{sol}(c)]}{p[Z_\mr{sol}(c)]}=
\begin{cases}
    \dfrac{1}{6Z_\mr{sol}\bigg(1-\dfrac{3.38}{\sqrt{Z_\mr{sol}}}+\dfrac{4.17}{Z_\mr{sol}}-\dfrac{1.55}{Z_\mr{sol}^{3/2}}\bigg)}, & \text{(DEMG)}\\
    \dfrac{1}{6Z_\mr{sol}\bigg(1-\dfrac{3.38}{\sqrt{Z_\mr{sol}}}+\dfrac{4.17}{Z_\mr{sol}}-\dfrac{1.55}{Z_\mr{sol}^{3/2}}\bigg)+1}, & \text{(Rolie-Poly)}\\
\end{cases}
\label{eqn:ratio_analytical}
\end{equation}
where the dimensionless apparent extensional relaxation time $\h{\lambda}_\mr{e}$ is expressed explicitly in Eq.~\ref{eqn:rel_time}, and the ratio of the disengagement time and the Rouse time $p(Z)$ is expressed in Eq.~\ref{eqn:Z}.

To substantiate the analytical prediction of Eq.~\ref{eqn:ratio_analytical}, we revisit the experimental data from two previous studies \cite{arnoldsCapillaryBreakupExtensional2010,
sachsenheimerExperimentalStudyCapillary2014} with two different polymer solutions at varying molecular weights and concentrations. To ensure the validity of the tube models, only measurements reported in the entangled regime ($c>c_\mr{e}$) are considered. The material properties and the experimental measurements of the selected systems are shown in Table~\ref{table:exp_res}. 
The molecular weight between entanglements $M_\mr{e}$ is evaluated at a temperature of \SI{20}{\celsius}, and the number of entanglements per chain for the corresponding polymer melts $Z_0$ is calculated based on the assumption of monodispersity.


\begin{table*}[!h]
\centering
\caption{Material properties, concentrations and experimentally measured ratios of the apparent shear and extensional relaxation times $\lambda_\mr{e}/\lambda_\mr{s}$ for the selected material systems from the previous studies.}
\begin{tabular*}{0.85\textwidth}{p{3cm}p{2.2cm}p{2cm}p{2cm}p{3cm}}
\hline
Materials & $M_\mr{e}$ (g/mol) & $M$ (g/mol) & $c$ (wt\%) & $\lambda_\mr{e}/\lambda_\mr{s}$\\
\hline
\multirow{4}{*}{PEO/Water \cite{arnoldsCapillaryBreakupExtensional2010}} & \multirow{4}{*}{$2200$ \cite{wypych2016handbook}} & \multirow{2}{*}{$1\times 10^6$} & $2.5$ & $0.168\pm 0.042$ \\
& & & $3.0$ & $0.197\pm 0.049$ \\
\cline{3-5}
& & \multirow{2}{*}{$2\times 10^6$} & $1.5$ & $0.125\pm 0.031$ \\
& & & $2.0$ & $0.104\pm 0.026$ \\
\hline
\multirow{9}{*}{PEO/Water \cite{sachsenheimerExperimentalStudyCapillary2014}} & \multirow{9}{*}{$2200$} & \multirow{4}{*}{$1\times 10^6$} & $2.0$ & $0.34$ \\
& & & $2.5$ & $0.24$ \\
& & & $3.0$ & $0.24$ \\
& & & $3.5$ & $0.15$ \\
\cline{3-5}
& & \multirow{2}{*}{$2\times 10^6$} & $1.5$ & $0.12$ \\
& & & $2.0$ & $0.10$ \\
\cline{3-5}
& & \multirow{3}{*}{$4\times 10^6$} & $1.0$ & $0.10$ \\
& & & $1.5$ & $0.05$ \\
& & & $2.0$ & $0.02$ \\
\hline
\multirow{3}{*}{PS/DEP \cite{sachsenheimerExperimentalStudyCapillary2014}} & \multirow{3}{*}{$16600$ \cite{mark2007physical}} & \multirow{3}{*}{$3\times 10^6$} & $3.0$ & $0.71$ \\
& & & $4.0$ & $0.50$ \\
& & & $5.0$ & $0.24$ \\
\hline
\multirow{2}{*}{PS/DEP \cite{clasenCapillaryBreakupExtensional2010}} & \multirow{2}{*}{$16600$} & \multirow{2}{*}{$13.2\times 10^6$} & $1.41$ & $0.36$ \\
& & & $1.77$ & $0.31$ \\
\hline
\end{tabular*}
\label{table:exp_res}
\end{table*}

\begin{figure}[!h]
    \centering
	\includegraphics[width=1\textwidth]{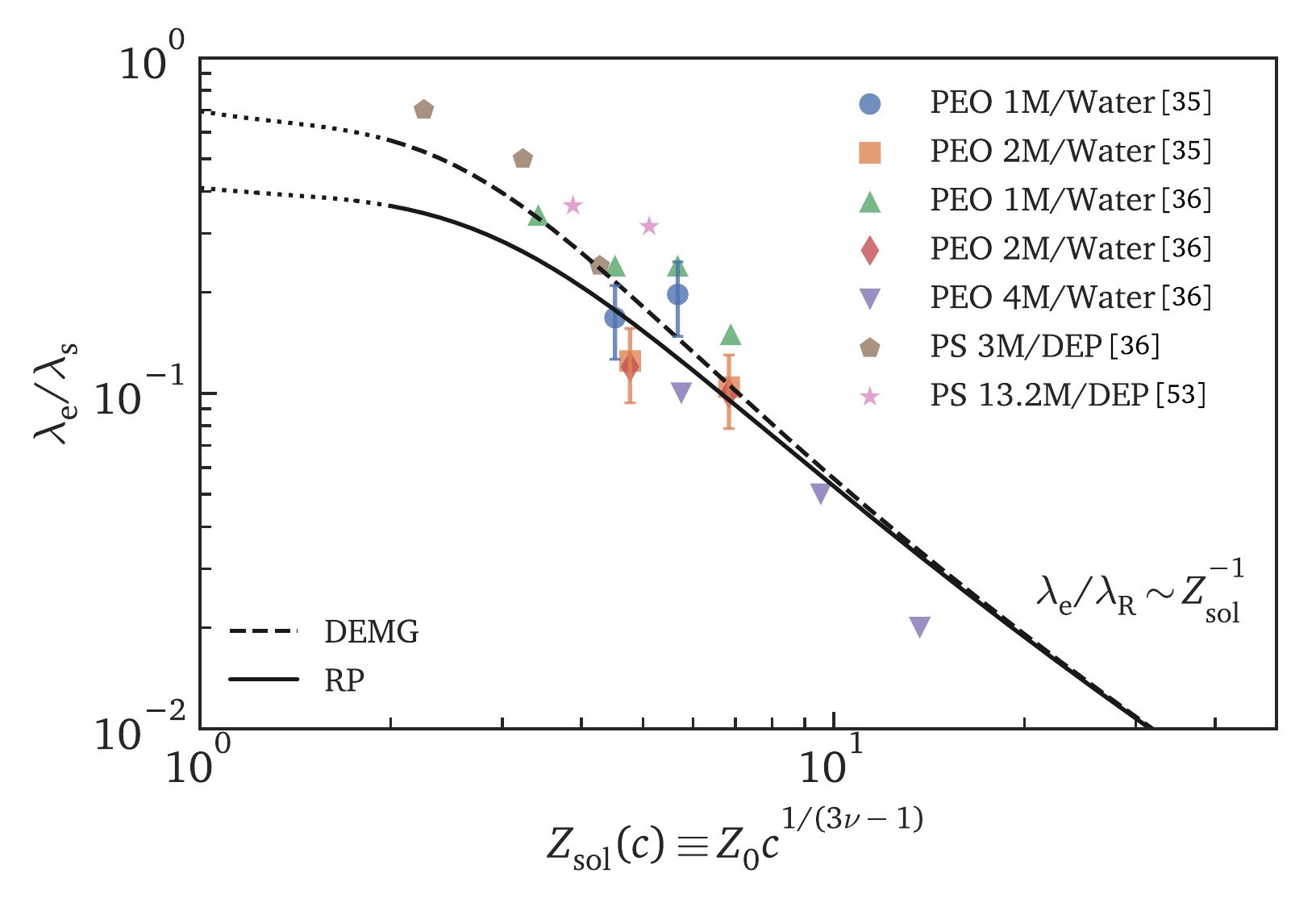}
    \caption{The ratio of apparent extensional and shear relaxation times $\lambda_\mr{e}/\lambda_\mr{s}$ against the number of entanglements in entangled polymer solutions $Z_\mr{sol}(c)=Z_0 ^{1/(3\nu-1)}$. The dashed and solid lines correspond to the prediction lines from the DEMG and the Rolie-Poly models, respectively, within the range of $2<Z_\mr{sol}(c)<100$. The experimental data for varying polymers, molecular weights and concentrations of the selected material systems broadly collapse onto a monotonically descending master curve, and are consistent with the predictions from both the selected tube models.}
    \label{fig:exp}
\end{figure}

We compare the experimental data reported in Table~\ref{table:exp_res} with the analytical prediction of Eq.~\ref{eqn:ratio_analytical} for the two tube models (dashed line: DEMG model; solid line: RP model) in Fig.~\ref{fig:exp}. Both models predict an overall descending trend for the ratio of the extensioanl and shear relaxation times with increasing number of entanglements per chain. Both predictions are applicable for entangled polymer solutions within the range of $2<Z_\mr{sol}(c)<100$ due to the validity of Eq.~\ref{eqn:Z} \cite{likhtmanQuantitativeTheoryLinear2002}. As $Z_\mr{sol}(c)$ increases, the two predictions coincide and the ratio of relaxation times asymptotically approaches a scaling relation of $\lambda_\mr{e}/\lambda_\mr{s}\sim Z_\mr{sol}^{-1}$. From Fig.~\ref{fig:exp}, it is clear that the experimental data of varying polymers, molecular weights and concentrations are broadly characterized by a single decreasing trend, which agrees well with the predictions from the two selected tube models without any additional fitting parameters. This overall consistency between the experimental data and the analytical prediction lines provides a physical insight into the origin of the substantial differences in the shear and extensional relaxation times reported in earlier studies: The tube networks constitute the primary topology of the entangled polymer systems and effectively shield the contribution to the shear rheology from a single polymer chain at small shear strains. However, the integrity of these tube structures becomes increasingly susceptible to a strong extensional flow as the tubes are forced to align. During this process, the free energy of each polymer chain decreases, and the material relaxation becomes progressively independent of the tube structures that are manifested at a larger length scale. 

\subsection{Extensional-thinning for highly entangled polymer systems}
\begin{figure}[!]
    \centering
    \includegraphics[width=\textwidth]{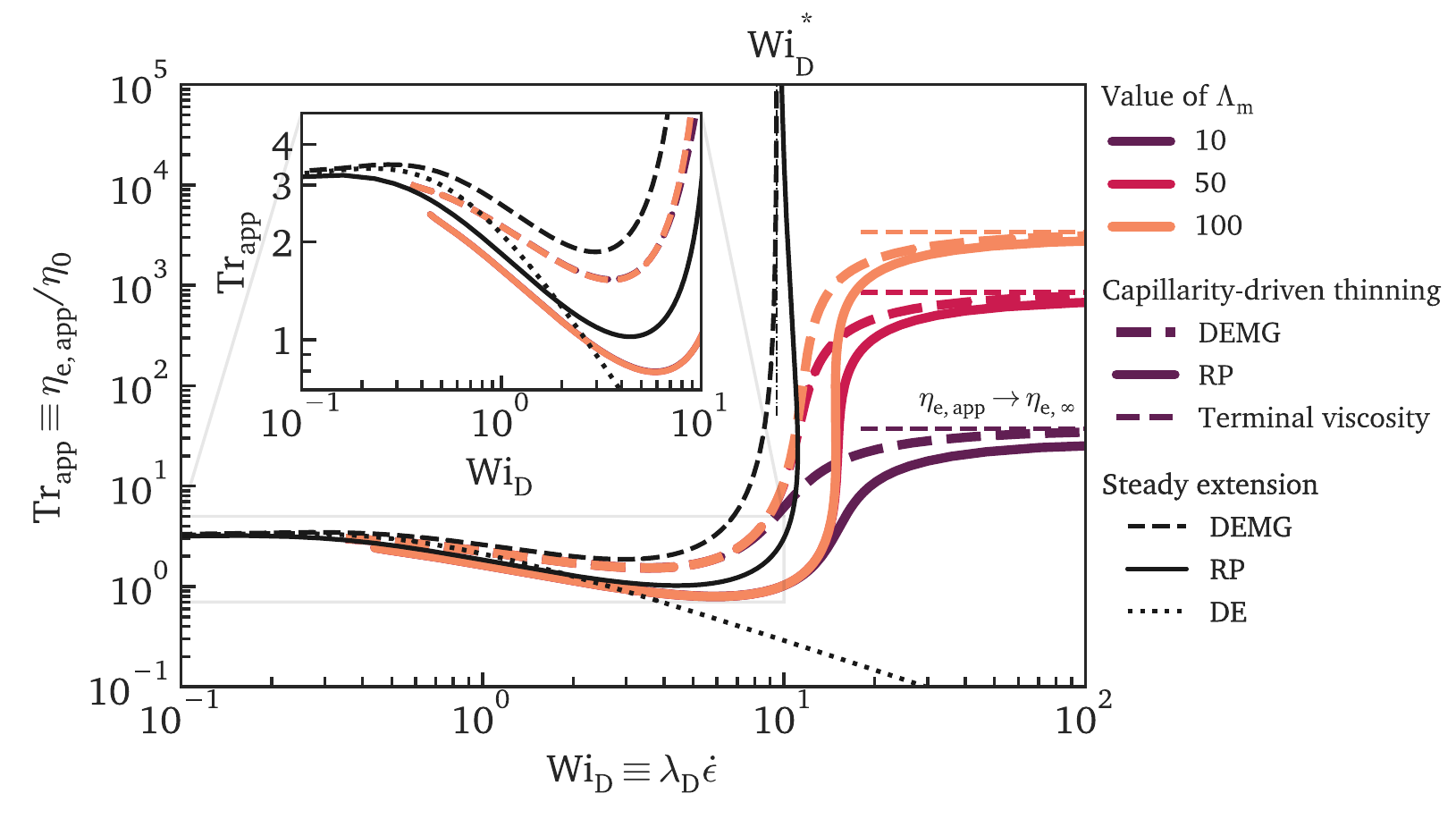}
    \caption{The apparent transient Trouton ratio $\mr{Tr}_\mr{app}=\eta_\mr{e,app}/\eta_0=\eta_\mr{e,app}/(G_\mr{N}\lambda_\mr{D})$ against the Weissenberg number $\mr{Wi}_\mr{D}=\dot{\epsilon}\lambda_\mr{D}$ extracted from the capillarity-driven thinning dynamics of the Rolie-Poly model (thick solid lines) with varying finite extensibilities $\Lambda_\mr{m}=\numlist{10;50;100}$ at a fixed elasto-capillary number $\mr{Ec}_0=1$ and a fixed number of entanglements $Z=10$. The transient extensional viscosities in both models show rate-thinning behavior at small strain rates ($\mr{Wi}_\mr{D}\lesssim 10$) due to tube reorientation. As $\mr{Wi}_\mr{D}$ increases, the polymer chain stretch becomes dominant in the extensional rheology, such that $\mr{Tr}_\mr{app}$ rapidly increases until reaching the terminal extensional viscosity set by the finite extensibility according to Eq.~\ref{eqn:terminal_vis} (horizontal dashed lines). Two reference lines are drawn from the steady extension predicted by the RP model (thin black solid line) and the original Doi-Edwards model (thin black dotted line). An infinite extensibility is imposed to the polymer chain stretch for the RP model. The prediction from steady extension of the RP model with chain stretch is broadly consistent with those extracted from the capillarity-driven thinning in the rate-thinning region, except for a different critical $\mr{Wi}_\mr{D}^*$ when the extensional viscosity diverges. In contrary, the prediction from the DE model (which does not describe additional chain stretch) illustrates the extensional rate-thinning trend expected from tube reorientation.}
    \label{fig:ext_viscosity}
\end{figure}

Another important feature extracted from measurements of capillarity-driven thinning dynamics of the entangled polymer solutions is the rate-thinning extensional viscosity \cite{dinicPowerLawsDominate2020}, and such extensional thinning can be increasingly pronounced as the polymer concentration further increases \cite{larsonModelingRheologyPolymer2015}. This trend can be effectively captured by the selected tube models. In Fig.~\ref{fig:ext_viscosity}, we show the transient extensional viscosity that would be computed from the temporal evolution of the filament radius with the same parameter variation as in Fig.~\ref{fig:intermediate_thinning}. The Trouton ratio is defined as the (transient) extensional viscosity scaled by the zero-shear viscosity $\eta_0$. For both single-mode tube models studied in this work, this zero-shear viscosity can be analytically calculated as $\eta_0=G_\mr{N}\lambda_\mr{D}$. The strain rates on the abscissa are nondimensionalized by the disengagement time $\lambda_\mr{D}$ to accommodate the DE model in which polymer chain stretch is absent. When $\mr{Wi}_\mr{D}\equiv\lambda_\mr{D}\dot{\epsilon}\rightarrow 0$, the Trouton ratio approaches a constant of $\mr{Tr}_{0}=3$. 

The transient Trouton ratio extracted from the capillarity-driven thinning dynamics predicted by the two tube models exhibit a complex and non-monotonic rate-dependent behavior. As we have illustrated in the previous sections, when $1\ll \mr{Wi}_\mr{D} \ll p(Z)$, chain reptation driven by Brownian motion overcomes the stretching induced by extensional flows. As a result, the tubes are forced to align progressively towards the extensional direction. 
 This tube-alignment mechanism generally leads to a decreasing extensional viscosity with increasing strain rate due to increased chain mobility in a relatively straightened tube network. In contrary, in a sufficiently fast extensional flow where $\mr{Wi}_\mr{D} \gg p(Z)$, the finite rate of polymer chain retraction dominates the polymer conformation. Beyond this point, the tube has been completely reoriented and hence does not effectively contribute to the extensional rheology of the systems. Instead, the resistance arising from the polymer chain stretch acting as entropic springs adds up to the overall viscoelastic response and thus increases the extensional viscosity. Eventually, as the chains reach full extension, the extensional viscosity reaches a terminal value given by Eq.~\ref{eqn:terminal_vis}. 

The transient apparent extensional viscosity $\eta_\mr{e,app}$ extracted from the time-dependent capillarity-thinning dynamics can be further compared with the response predicted from a steady homogeneous extensional flow. As shown in Fig.~\ref{fig:ext_viscosity}, the steady extensional viscosity from the RP model (black solid line) is broadly consistent with the transient predictions (thick colored lines). The rate of extensional-thinning predicted by the DEMG model (not displayed in the figure) is subtly distinct due to the additional CCR effect. A major discrepancy between the steady and transient responses is the critical Weissenberg number $\mr{Wi}_\mr{D}^*$ where the extensional viscosity diverges. From analytical calculations, the steady extension predicts that $\mr{Wi}_\mr{D}^*=p(Z)+1/2$, while for the transient viscosity, a delayed response is identified as $\mr{Wi}_\mr{D}^*=4p(Z)/3$. This increased front factor can be attributed to a non-vanishing rate of chain stretch $\Lambda$ during the temporal evolution of the filament radius.  

In practice, if the rate-thinning behavior is prominent in the capillarity-driven thinning dynamics for a tested fluid, the elasto-capillary number is reasonably large. As a result, the exponential-thinning regime of the filament radius derived from the polymer chain stretch is unlikely to be observed within the measurable range. To parsimoniously describe the rate-thinning kinematics, we numerically calculate the original Doi-Edwards (DE) model with no polymer chain stretch. A differential form of the constitutive equation is taken from Eq.~\ref{eqn:demg_model} by setting $\Lambda=1$. In this scenario, the magnitude of reorientation $\Delta S$ has an analytical solution, and a modified version of the DE model with the addition of the solvent viscosity can be obtained to fit the experimental data without approaching singularity at high strain rates.

\begin{figure}[!]
    \centering
    \includegraphics[width=\textwidth]{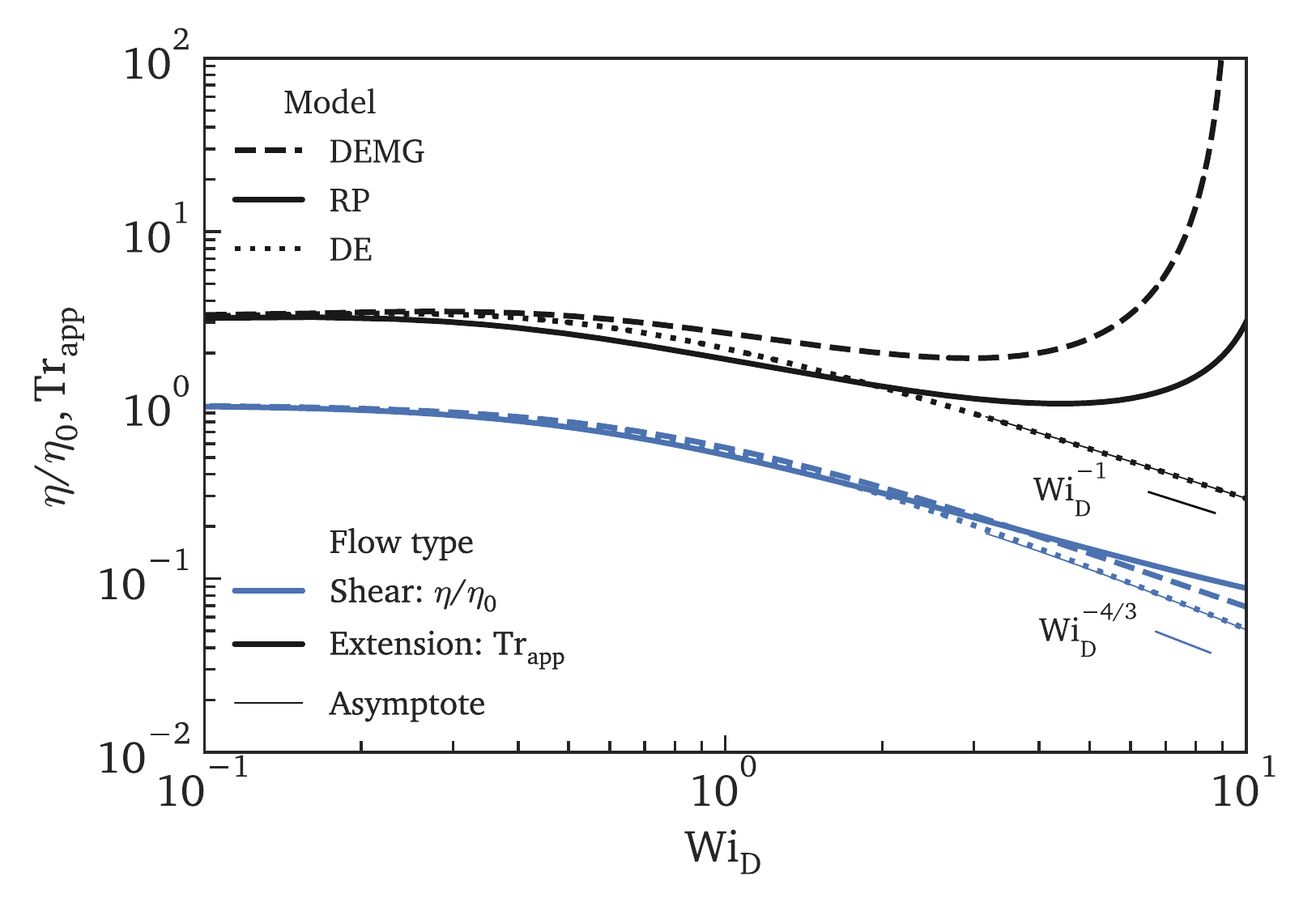}
    \caption{Dimensionless shear and extensional viscosities, $\eta/\eta_0$ (blue lines) and $\mr{Tr}_\mr{app}$ (black lines) against the Weissenberg number $\mr{Wi}_\mr{D}$ for the Rolie-Poly model (solid line), the DEMG model (dashed line) and the Doi-Edwards model (dotted line) at $\mr{Ec}_0=1$, $Z=10$ and $\Lambda_\mr{m}\rightarrow\infty$ if applicable. The shear viscosities among the three tube models shows a consistent shear-thinning trend until close to $\mr{Wi}_\mr{D}=10$. In extensional flow, a broadly consistent extensional-thinning trend is observed at low $\mr{Wi}_\mr{D}$, followed by distinctly different trend for $\mr{Wi}_\mr{D}\gg 1$. The two tube models with the additional polymer chain stretch predict a rapidly increasing extensional viscosity, while the DE model predicts a steadily decreasing trend. The asymptotic solutions of both the shear and extensional viscosities for the DE model at high $\mr{Wi}_\mr{D}$ are plotted as thin lines, which exhibit a power-law scaling of $\mr{Wi}_\mr{D}^{-4/3}$ and $\mr{Wi}_\mr{D}^{-1}$, respectively.}
    \label{fig:ext_viscosity_model}
\end{figure}

Fig.~\ref{fig:ext_viscosity_model} shows the dimensionless shear (blue) and extensional (black) viscosities of the RP model (solid line), the DEMG model(dashed line) and the Doi-Edwards model (dotted line) at $\mr{Ec}_0=1$, $Z=10$ and $\Lambda_\mr{m}\rightarrow\infty$ (if applicable). A range of $0.1<\mr{Wi}_\mr{D}<10$ is observed to be consistent with the extensional-thinning region from the capillarity-driven thinning dynamics. All three prediction lines of the shear viscosity show a consistent shear-thinning trend, which is resulted from the tube reorientation. The extensional viscosity, on the contrary, shows a diverse trend between the two tube models with the effect of the polymer chain stretch, and the DE model. Nevertheless, all three curves exhibit a similar scaling relation in most of the extensional-thinning regime. The asymptotic solutions of the shear and extensional viscosities as $\mr{Wi}_\mr{D}\rightarrow \infty$ from the DE model are representative of the rate-dependent material response from the tube reorientation. They are analytically calculated up to the second dominant order as 
\begin{subequations}
\begin{align}
\dfrac{\eta}{G\lambda_\mr{D}}&=\dfrac{3}{12^{1/3}\mr{Wi}_\mr{D}^{4/3}}\dfrac{1}{1-\dfrac{4}{12^{2/3}\mr{Wi}_\mr{D}^{2/3}}}\sim \mr{Wi}_\mr{D}^{-4/3}, \\
\mr{Tr}_\mr{D}&=\dfrac{9}{3\mr{Wi}_\mr{D}+1}\sim \mr{Wi}_\mr{D}^{-1}.
\end{align}
\end{subequations}
Therefore, the shear and extensional viscosities asymptotically approach a scaling law of $\mr{Wi}_\mr{D}^{-4/3}$ and $\mr{Wi}_\mr{D}^{-1}$, respectively, as $\mr{Wi}_\mr{D}$ increases. The larger-than-unity power in the scaling from the shear flow is well-known to produce flow instability \cite{dealyMolecularStructureRheology2006}. Nevertheless, the linearly inverse extensional-thinning trend predicted by the DE model shows a pronounced strain-rate dependence and provides a feasible explanation to the extensional rheology of a wider range of entangled polymer solutions. 

\subsection{Demonstration: Polyethylene oxide}
We choose a specific polymer solution, polyethylene oxide (PEO) at a molecular weight of 1~MDa to demonstrate the evolution of the capillarity-driven thinning dynamics over a wide range of the concentrations covering both the dilute and the entangled regimes. Table~\ref{table:peo} shows the properties and configurations of the aqueous PEO solutions.
\begin{table*}[!h]
\centering
\caption{Properties and configurations of the aqueous PEO solutions applied to the numerical calculation of the capillarity-thinning dynamics at varying concentrations.}
\begin{threeparttable}
\begin{tabular*}{\textwidth}{p{12cm}p{3cm}}
\hline
\textbf{Properties} & \textbf{Values}\\
\hline
Molecular weight $M$ (\si{\gram\per\mole}) & $10^6$ \\
Monomer size $M_0$ (\si{\gram\per\mole})   & $44$   \\
Degree of polymerization $N$ & 22727\tnotex{tnote:dop} \\
Molecular weight between entanglements in melts $M_\mr{e}$ (\si{\gram\per\mole}) & 2200 \cite{wypych2016handbook} \\
Characteristic ratio $C_{\infty}$ & 5.6 \cite{hiemenz2007polymer}\\
Monomer length $l$ (\si{\nm}) & 0.35  \cite{hiemenz2007polymer} \\
Kuhn step $N_K$  & 4058\tnotex{tnote:ks} \\
Kuhn length $b_K$ (\si{\nm}) & 2\tnotex{tnote:kl} \\
End-to-end distance $\left<R_0^2\right>^{1/2}$ (\si{\nm}) & 125\tnotex{tnote:ete} \\
Statistical length $b$ (\si{\nm}) & 0.83\tnotex{tnote:sl} \\
Temperature for solutions $T$ (\si{\celsius}) & 25 \\ 
Surface tension $\Gamma$ (\si{\mN\per\m}) & 62.2 \cite{caoMolecularWeightDependence1994} \\
Initial filament radius $R_0$ (\si{\mm}) & 1 \\
Zimm time $\lambda_\mr{Z}$ (\si{\ms}) & 0.51 \cite{tirtaatmadja2006drop} \\
Friction coefficient of monomer $\zeta_0$ & $1.4\times 10^{-11}$ \\
Rouse time $\lambda_\mr{R}$ (\si{\ms}) & 20.2\tnotex{tnote:rt} \\
Overlap concentration $c^*$ (\si{\wtpc}) & 0.161 \cite{tirtaatmadja2006drop} \\
Entanglement concentration $c_\mr{e}$ (\si{\wtpc}) & 1.7 \cite{sachsenheimerExperimentalStudyCapillary2014} \\
\hline
\end{tabular*}
\bigskip
\begin{tablenotes}
	\item\label{tnote:dop} $N=M/M_0$.
	\item\label{tnote:ks} $N_K=N/C_{\infty}$ and $N_K=\Lambda_\mr{m}^2$ in dilute polymer solutions \cite{hiemenz2007polymer}.
	\item\label{tnote:kl} $b_K=C_{\infty} l$ \cite{hiemenz2007polymer}.
	\item\label{tnote:ete} $\left<R_0^2\right>^{1/2}=\sqrt{N_K}b_K$ \cite{hiemenz2007polymer}.
	\item\label{tnote:sl} $b=b_K\sqrt{N}$ \cite{dealyMolecularStructureRheology2006}.
	\item\label{tnote:rt} $\lambda_\mr{R}=\zeta_0 N^2 b^2/(6\pi^2 kT)$ \cite{dealyMolecularStructureRheology2006}.
\end{tablenotes}
\end{threeparttable}
\label{table:peo}
\end{table*}
The modulus in the FENE-P for the dilute polymer solutions can be calculated from the kinetic theories as $G=nKT=ckT/\left<R_0^2\right>^{3/2}$ \cite{wagnerAnalyticSolutionCapillary2015}. The plateau modulus of the entangled polymer solutions can be calculated from the Graessley-Fetters definition \cite{dealyMolecularStructureRheology2006} as
\begin{equation}
G_\mr{N}(c)=\dfrac{4\rho_\mr{sol} c RT}{5M_\mr{e}(c)},
\end{equation}
where $\rho_\mr{sol}$ is the solution density, and the molecular weight between entanglements for solutions $M_\mr{e}(c)$ is expressed in Eq.~\ref{eqn:me_sol}. The constitutive equation for semi-dilute polymer solutions is taken from Prabhakar et al. \cite{prabhakarInfluenceStretchingInduced2016} by adding a concentration- and strain-dependent correction factor to the relaxation time in the FENE-P model as
\begin{equation}
\lambda(c/c^*,\Lambda)=\nu(c/c^*, \Lambda)\lambda_\mr{Z},
\end{equation}
where $c^*$ is the overlap concentration, and $\lambda_\mr{Z}$ is the Zimm time that describes the relaxation time in the regime of infinite dilution. The polymer chain stretch $\Lambda$ is expressed \cite{Hinch1973,
birdDynamicsPolymericLiquids1987,
wagnerAnalyticSolutionCapillary2015,
prabhakarInfluenceStretchingInduced2016} as
\begin{equation}
\Lambda=\dfrac{\sqrt{\Tr\left<\boldsymbol{Q}\boldsymbol{Q}\right>}}{\left<R_0^2\right>^{1/2}},
\end{equation}
where $\boldsymbol{Q}$ is the end-to-end vector of a single polymer molecule, and the angle brackets correspond to the ensemble average. 

To illustrate the filament thinning profiles, six concentrations are selected in the dilute and semi-dilute regimes ($c=\SIlist{0.0016;0.081;0.24}{\wtpc}$; predicted by the FENE-P model with a corrected relaxation time; marked in blue) as well as the entangled regime ($c=\SIlist{2.0;2.5;3.0}{\wtpc}$; predicted by the RP model; marked in yellow and red). Based on the properties listed in Table~\ref{table:peo}, we further calculate the concentration-specific parameters that are used for the numerical calculation as shown in Table~\ref{table:peo_sol}. It is noted that both the (plateau) moduli and the number of entanglements increase with the concentration, while the maximum polymer chain stretch decreases as the polymers are increasingly entangled, thus shortening the length of one tube segment. 
\begin{table*}
\centering
\caption{Selected concentrations of the aqueous PEO solutions to illustrate the filament thinning dynamics with concentration-specific properties. The colored lines next to the concentrations are consistent with Fig.~\ref{fig:PEO_demo}.}
\begin{tabular*}{\textwidth}{p{4cm}p{2cm}p{2cm}p{3cm}p{2cm}p{2cm}}
\hline
Concentrations (\si{\wtpc}) & $c/c^*$ & $c/c_\mr{e}$ & $G$ or $G_\mr{N}$ (Pa) & $Z_\mr{sol}$ & $\Lambda_\mr{m}^2$\\
\hline
0.0016 (\drawlinesc[color=sol1, line width=2pt]) & $9.9\times 10^{-3}$ & $9.4\times 10^{-4}$ & 0.02 & N/A& 4058\\
0.081 (\drawlinesc[color=sol2, line width=2pt]) & 0.50 & 0.048 & 1.06 & N/A & 4058\\
0.24 (\drawlinesc[color=sol3, line width=2pt]) & 1.49 & 0.14 & 3.17 & N/A & 4058\\
\hline
2.0 (\drawlinesc[color=sol4, line width=2pt]) & 12.42 & 1.17 & 135.59 & 1.5 & 1187\\
2.5 (\drawlinesc[color=sol5, line width=2pt]) & 15.53 & 1.47 & 224.02 & 2.3 & 898\\
3.0 (\drawlinesc[color=sol6, line width=2pt]) & 18.63 & 1.76 & 337.63 & 3.4 & 715\\
\hline
\end{tabular*}
\label{table:peo_sol}
\end{table*}

\begin{figure}[!]
    \centering
    \includegraphics[width=\textwidth]{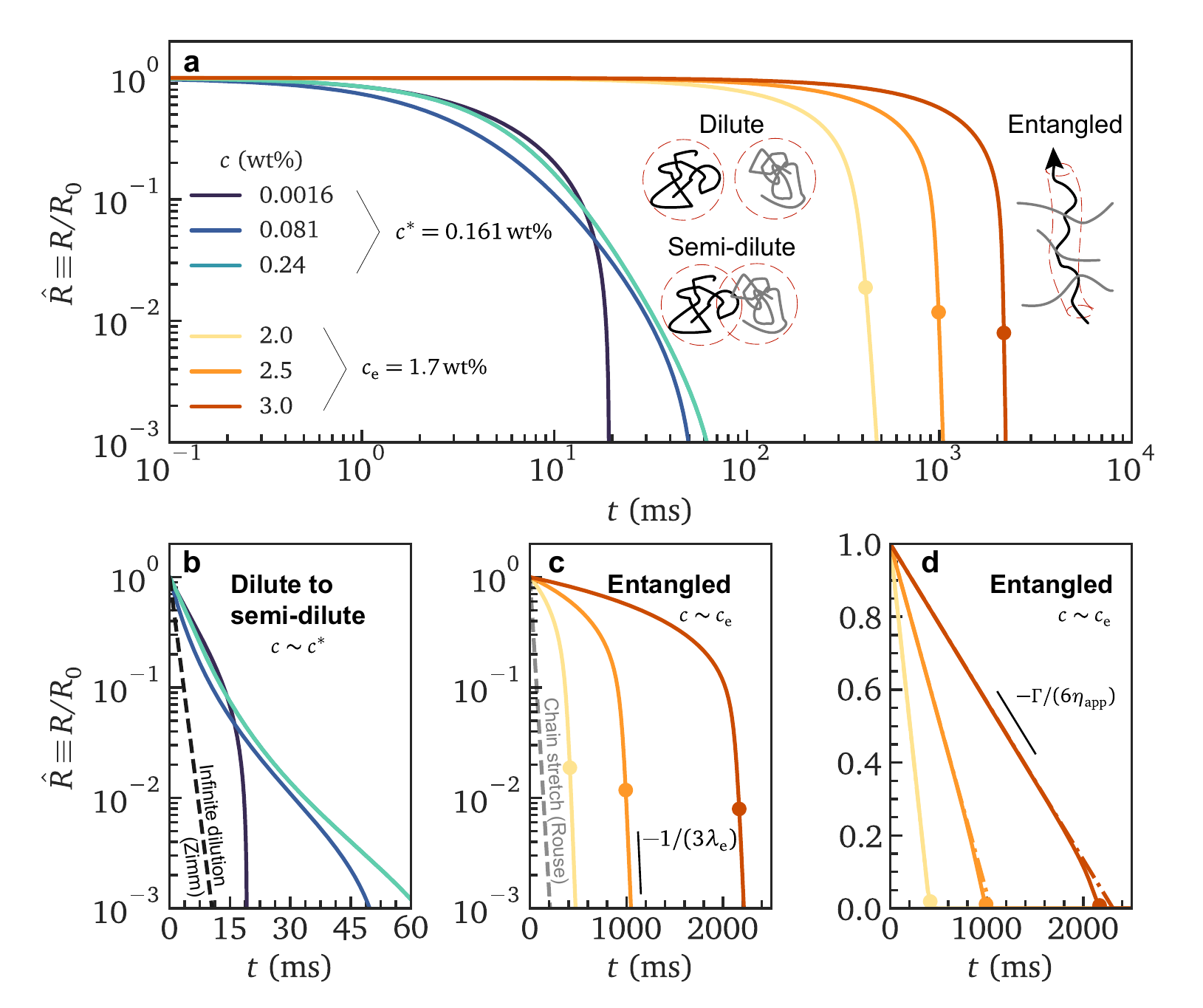}
    \caption{Numerical calculations of the temporal evolution of the filament radius for aqueous polyethylene oxide (PEO) solutions over the dilute and entangled concentrations. The time axis is dimensional, and the filament radius is nondimensionalized by the initial radius specified in Table~\ref{table:peo}. (a) The temporal evolution of the filament radius in all concentration regimes with both axes plotted in logarithmic scales. The filament thinning is progressively retarded as the concentration increases, when the initial slow-thinning becomes more evident. (b) The temporal evolution of the filament radius in the dilute and semi-dilute regimes described by the FENE-P model with a corrected relaxation time. Only the ordinate is plotted in a logarithmic scale. The asymptote using the Zimm time ($\lambda_\mr{e}=\lambda_\mr{Z}$) is plotted in black dashed line to show the filament thinning in the limit of infinite dilution. (c) The temporal evolution of the filament radius in the entangled regime described by the Rolie-Poly model. Only the ordinate is plotted in a logarithmic scale. The asymptote using the Rouse time ($\lambda_\mr{e}=\lambda_\mr{R}/2$) is plotted in gray dashed line to show the filament thinning in the limit of no tube reorientation. (d) The temporal evolution of the filament radius in the entangled regime described by the Rolie-Poly model plotted in a linear scale. The dashed dotted lines show the predictions from a Newtonian fluid with the apparent zero-shear viscosity from Eq.~\ref{eqn:eta0}. The markers in (a), (c) and (d) for the entangled solutions indicate the transition of the filament thinning to an exponential-thinning trend under an elasto-capillary balance.}
    \label{fig:PEO_demo}
\end{figure}

In Fig.~\ref{fig:PEO_demo}, the time axis is dimensional to provide an estimate of the breakup time. As the concentration increases from below $c^*$ to above $c_\mr{e}$, the filament thinning is progressively retarded by several orders of magnitude. In Fig.~\ref{fig:PEO_demo}(b) and (c), we show the two exponential-thinning asymptotes using the Zimm time (black dashed lines, $\lambda_\mr{e}=\lambda_\mr{Z}$) and the Rouse time (gray dashed line, $\lambda_\mr{e}=\lambda_\mr{R}/2$ from Table~\ref{table:SAOS}), respectively. These two asymptotes set two lower bounds for the temporal evolution of the filament radius for dilute and entangled polymer solutions, and can be practically used to identify the most appropriate constitutive model to extract accurate extensional rheological properties. The kinematics of the filament radius evolve from an exponential-thinning or slightly-deviated exponential-thinning trend under the elasto-capillary balance to a smoother thinning trend at concentrations above $c_\mr{e}$. In this stage, the initial slow thinning becomes more prominent as the highly entangled tube structures increasingly impede the disengagement of the polymer chains in extensional flow. In practice, the measurement of the filament radius is limited by the resolution of the image capturing devices. A measurable range that covers three orders of magnitude down to $R/R_0<0.001$ is less accessible. As a result, the slow-thinning trend that occurs at the beginning of the filament thinning is more evident in the experimental data. To show this, we mark the time $t_\mr{V\mbox{-}E}$ when the transient strain rate of the filament reaches the critical strain rate under an elasto-capillary balance at $\mr{Wi}=2/(3\h{\lambda}_\mr{e})$ in the entangled regime (circles). Beyond this point, the filament radius decays in an exponential trend until the polymer chain stretch approaches the maximum stretch, as illustrated in Fig.~\ref{fig:intermediate_thinning}. As shown in Fig.~\ref{fig:PEO_demo}(d), where the filament radius is plotted in a linear scale, the transition to the exponential thinning occurs close to the filament pinch-off. The majority of the filament thinning profile agrees well with the prediction from a Newtonian fluid, of which the shear viscosity is identical to the apparent zero-shear viscosity from Eq.~\ref{eqn:eta0}.

\begin{figure}[!]
    \centering
    \includegraphics[width=\textwidth]{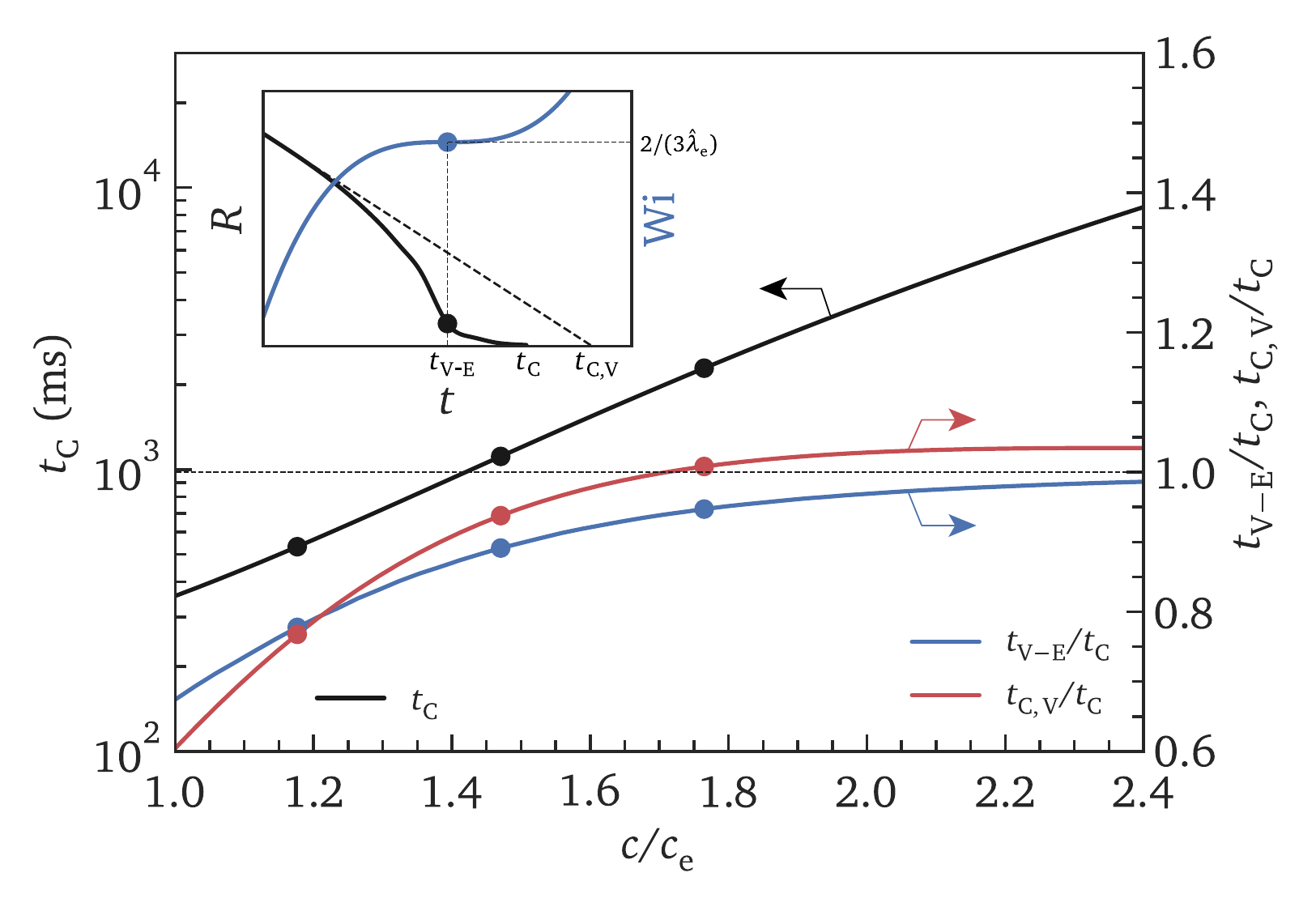}
    \caption{The dimensional filament breakup time $t_\mr{C}$ (black line) by extrapolating the prediction line from the RP model to $R=0$. Another two timescales $t_\mr{V\mbox{-}E}$ (blue) and $t_\mr{C,V}$ (red) are plotted by taking the ratio to $t_\mr{C}$. The three selected concentrations in the entangled regime illustrated in Fig.~\ref{fig:PEO_demo} are marked.}
    \label{fig:PEO_timescale}
\end{figure}

From the numerical calculation of the filament thinning profiles in the entangled regime, we can quantify how the filament thinning dynamics evolve with the concentration by inspecting three timescales: (1) the transition time $t_\mr{V\mbox{-}E}$ when the exponential-thinning trend is manifested at $\mr{Wi}=2/(3\h{\lambda}_\mr{e})$, following visco-capillary thinning behavior; (2) the filament breakup time $t_\mr{C}$ obtained by extrapolating the prediction line from the RP model to $R=0$; (3) the filament breakup time $t_\mr{C,V}$ predicted by the Newtonian fluid with the apparent zero-shear viscosity in Eq.~\ref{eqn:eta0}. As shown in Fig.~\ref{fig:PEO_timescale}, we observe that the filament breakup time $t_\mr{C}$ increases in a broadly exponential trend with the concentration above $c_\mr{e}$. At $c/c_\mr{e}\approx 1.7$, the filament breakup time $t_\mr{C}$ is approximately identical to $t_\mr{C,V}$, where the rate-thinning effect induced by the tube reorientation cancels off with the rate-thickening effect from the polymer chain stretch. Beyond this point, the elasto-capillary transition time $t_\mr{V\mbox{-}E}$ and the Newtonian breakup time $t_\mr{C,V}$ defined previously become increasingly close to $t_\mr{C}$, while $t_\mr{C,V}>t_\mr{C}>t_\mr{V\mbox{-}E}$. The evolution of the three timescales illustrates a delayed transition to the exponential thinning regime under an elasto-capillary balance as the concentration increases. At sufficiently large concentrations, the tube reorientation that induces a rate-thinning behavior is increasingly critical in determining the final breakup time than the polymer chain stretch where the extensional viscosity increases with the strain rate. Therefore, we can identify the key features in the complex capillarity-driven thinning dynamics of the entangled polymer solutions more accurately. 

\section{Conclusion}
We numerically calculate the capillarity-driven thinning dynamics of the entangled polymer systems by the application of two selected tube models: the Doi-Edwards-Marrucci-Grizzuti (DEMG) model and the Rolie-Poly (RP) model. Both models incorporate the interaction of polymer chains through reptative motion, as well as the finite rates of retraction of a single polymer molecule. The RP model adds another non-reptative mechanism of the convective constraint release (CCR) effect. The two models are substituted in the stress balance equation in the filament thinning configuration assuming cylindrical filament shapes and no fluid inertia. By inspecting the kinematics of the resulting filament thinning profiles, we find three distinct regimes of the dynamics that are dominated by different constitutive parameters.

At the beginning of the filament thinning, the tube networks are forced to reorient under a moderate extensional rate, and the polymer chains within the tube remain unstretched. This mechanism results in slowed filament thinning with a rate-thinning trend in the extensional viscosity. An analytical solution for the rate of the shear- and extensional-thinning triggered by the reptation can be obtained from the Doi-Edwards model. As the filament thinning progresses, the stretch of the polymer chain stretch becomes increasingly dominant in the extensional rheology. As a result, the temporal evolution of the filament radius asymptotically approaches an exponential-thinning decay, similar to the filament thinning trend predicted by the FENE-P model under the elasto-capillary balance for dilute polymer solutions. An apparent extensional relaxation time can be extracted from the exponential-thinning decay and is found to approach half the Rouse time when the number of entanglements per chain grows sufficiently large. This extensional relaxation time is notably smaller than the shear relaxation time, \textit{i.e.}, the disengagement time. This finding contradicts with the predictions by the constitutive models that characterize dilute polymer solutions such as the FENE-P model, in which the measured relaxation times are independent of the flow types. Such discrepancy in the relaxation times reveals a different mechanism in the entangled polymer solutions that predominates in the polymer relaxation in a strong extensional flow. We further obtain an analytical expression for the ratio of the shear- and extensional relaxation times that is free of additional fitting parameters. This analytical result is substantiated by a number of different entangled polymer solutions from the previous studies. For both the selected tube models, as the extensional strain accumulates and approaches the finite extensibility of a single polymer chain close to the filament breakup, the temporal evolution of the filament radius deviates from the exponential trend. A constant terminal extensional viscosity can be obtained that incorporates the contribution from the polymer chain alignment, and its magnitude is much higher than the zero-rate viscosity. 

Finally, the capillarity-driven thinning dynamics of the aqueous polyethylene oxide solutions with a molecular weight of \SI{1}{\mega Da} are numerically calculated based on the real physical properties. A wide range of the concentrations over the dilute, semi-dilute and entangled regimes described by multiple constitutive models are compared in the same plot. As the concentration increases, the filament breakup is progressively retarded, and the filament thinning kinematics deviate from the well-studied exponential-thinning trend under the elasto-capillary balance predicted by the FENE-P model. The rate-thinning behavior that is prominent in a highly elastic entangled polymer system becomes increasingly evident within the measurable range. We expect that the new capillarity-driven thinning dynamics predicted by the tube models presented in this work will lead to more accurate extensional rheological characterizations of the entangled polymer solutions. We hope that new insights can be added to better understand and optimize a number of industrial applications that involve the processing of these complex polymer systems.

\begin{acknowledgement}
    J.D. and G.H.M. would like to thank Ford Motor Company for the financial support on this project.
\end{acknowledgement}



\bibliography{roliePoly}

\end{document}